\documentclass[a4paper,fleqn,usenatbib]{mnras}

\usepackage{newtxtext,newtxmath}
\usepackage{xcolor}

\usepackage[T1]{fontenc}
\usepackage{ae,aecompl}

\usepackage{graphicx}	
\usepackage{amsmath}	

\title[AGN powered SMG]{The nature of sub-millimetre galaxies I: A comparison of AGN and star-forming galaxy SED fits}

\author[T. Shanks et al.]{
T. Shanks,\thanks{E-mail: tom.shanks@durham.ac.uk (TS)}$^1$
B. Ansarinejad$^1$,
R.M. Bielby$^1$,
I. Heywood$^2$, 
N. Metcalfe$^1$,
L. Wang$^3$.\\
$^1$CEA, Physics Department, Durham University, South Road, Durham, DH1 3LE, UK\\
$^2$Astrophysics, Department of Physics, University of Oxford, Keble Road, Oxford, OX1 3RH, UK\\
$^3$SRON Netherlands Institute for Space Research, Landleven 12, 9747 AD, Groningen, The Netherlands
}

\date{Accepted --. Received --; in original form 2020 October 14}

\pubyear{2020}

\begin{document}
\label{firstpage}
\pagerange{\pageref{firstpage}--\pageref{lastpage}}
\maketitle

\begin{abstract}
High redshift sub-millimetre galaxies (SMGs) are usually assumed to be
powered by star-formation. However, it has been clear for some time that
$>$20\% of such sources brighter than $\approx3$mJy host quasars. Here
we analyse a complete sample of 12 sub-mm LABOCA/ALMA 870 $\mu$m sources
in the centre of the William Herschel Deep Field (WHDF) with
multi-wavelength data available from the X-ray to the radio bands.
Previously, two sources were identified as X-ray absorbed quasars at
$z=1.32$ and $z=2.12$. By comparing their spectral energy distributions
(SEDs) with unabsorbed quasars in the same field, we confirm that they
are dust reddened although at a level significantly lower than implied
by their X-ray absorption. Then we compare the SED's of all the sources
to dust-reddened AGN and star-forming galaxy models. This optical/NIR
comparison combined with Spitzer MIR colours and faint Chandra X-ray
detections shows that 7/12 SMGs are best fitted with an obscured quasar
model, a further 3/12 show no preference between AGN and star-forming
templates, leaving only a $z=0.046$ spiral galaxy and one unidentified
source. So in our complete sample, the majority (10/12) of  bright SMGs
are at least as likely to fit an AGN  as a star-forming galaxy template,
although no claim is made to rule out the latter as SMG power sources.
We then suggest modifications to a previous SMG number count model and
conclude that obscured AGN in SMGs may still provide the dominant
contribution to both the hard X-ray and sub-millimetre backgrounds.

\end{abstract}




\section{Introduction}
\label{intro}

It is now more than two decades since the first blank field surveys
using the Submillimetre Common-User Bolometer Array (SCUBA) on the James
Clerk Maxwell Telescope (JCMT) revealed the existence of large numbers
of highly luminous Far Infra-Red (FIR) sources, usually referred to as
sub-millimetre galaxies or SMGs \citep{Smail97, Barger98}. These objects
were quickly found to be at high redshift  and to be heavily obscured by
dust (e.g. \citealt{Blain02}). But the identity of luminous sub-mm
sources is still controversial. The standard view is that they are
hyper-luminous starburst galaxies, seen in an obscured phase of their
evolution (e.g. \citealt{Alexander05}). In this interpretation they may
be involved in the  origin of early-type galaxies. But because of their
high star-formation rate (SFR) and mass they then present a problem for
the standard cosmology that still has to invoke a `top-heavy' stellar
Initial Mass Function (IMF) in starbursts to explain them
(\citealt{Baugh05, Lacey08}). Bars and even spiral arms have been
detected in ALMA (Atacama Large Millimeter/submillimeter Array) images
and claimed as support for the star-forming galaxy picture
\citep{Hodge16, Gullberg19}. However, these features are always only
seen on scales an order of magnitude smaller than seen in local
classical spiral galaxies.

An alternative hypothesis is that the bright sub-millimetre population
is mainly powered by Active Galactic Nuclei (AGN). Many sub-mm sources
contain AGN but the usual view is that they are sub-dominant to
star-formation in powering these sources (e.g. \citealt{Laird10}). But
there are strong arguments for considering obscured AGN as being
candidates for powering the SMG population. Obscured quasars are the
primary candidates to explain the ``missing'' hard X-ray background
(e.g. \citealt{Comastri95}). Since these highly absorbed sources are
likely to be dust-rich objects, they would be expected to have
substantial luminosities in the infrared where the reprocessed light is
emitted. Obscured AGN models have been previously shown to give a
reasonable fit to the bright end of the sub-millimetre  source counts,
while star-forming galaxies are expected to make the dominant
contribution at fainter fluxes (see Fig. 12 of \citealt{Hill11a}
following \citealt{Gunn99}). If the sub-millimetre  emission results
from a dust torus irradiated by an AGN instead of star-formation, it
must lie far enough ($\approx0.5$kpc) from the central engine to
maintain a cool ($\sim35$K) temperature and simple torus models confirm
feasibility (e.g. \citealt{Andreani99,Kur03,Siebenmorgen14}).
\cite{Hill11a, Hill11b} also showed that an obscured AGN model also
fitted the bimodality of the Herschel Spectral and Photometric Imaging
Receiver (SPIRE) $n(z)$, with the star-forming and AGN components
dominating at low and high  $z$ respectively, while predicting that
$\approx20$\% of bright SMGs should be X-ray sources even at soft
energies. 

There now exists  clear evidence of AGN activity in 20-40\% of bright
sub-millimetre  sources, confirming this  prediction. First, early
$1.''4$  ALMA imaging by \citet{Wang13} found that $17\pm6$\% of
Extended Chandra Deep Field South (ECDFS) SMG's are X-ray sources. More
recently, \citet{Cowie18} in their Super GOODs ALMA survey found
evidence for bright SMGs being dominated by hard X-ray sources while
fainter sub-mJy SMGs were dominated by soft X-ray sources more
consistent with star-forming galaxies. \cite{Franco18} found that at
least 40\% of bright SMGs host an X-ray AGN, compared to $\approx14$\%
for other galaxies of similar mass and redshift. Finally,
\citet{Stach19} find that $18\pm10$\% of bright SMGs are AGN as
identified by power-law SEDs and other diagnostics. These significant
AGN fractions are all the more remarkable given the $2-3\times$  more
near-Compton-thick AGN that may be wholly invisible even to $\approx10$
keV X-ray surveys. We also note that  AGN-powered SMG number count
models generally assume that fainter sub-mJy SMGs will be dominated by 
star-forming galaxies while  the $>$1mJy luminous SMG population will be
dominated by AGN powered sources, consistent with these recently
reported results. With the advent of high resolution ALMA observations,
submm source counterparts can be more accurately identified and then
tested for the presence of an active nucleus. In about $\approx20$\% of
cases, broad or high excitation emission lines can be seen. In the other
$\approx80$\%, we can now check how well a dust obscured AGN  Spectral
Energy Distribution (SED) fits the ALMA counterpart in the optical/IR.

Here, we shall test how well a  sample of 12 SMGs, complete to
$\approx3$ mJy at 870 $\mu$m, can be fitted with dust obscured AGN
template SEDs. They were  originally identified in an $11.'4$ diameter
area of the William Herschel Deep Field (WHDF, \citealt{NM96,NM01,NM06}) by
\cite{Bielby12} using the Large APEX Bolometer Camera (LABOCA) on the
Atacama Pathfinder Experiment (APEX) telescope. We shall compare the
quality of these fits to those for similarly absorbed star-forming
galaxies. We are helped by having new, high resolution, high signal-noise (S/N) ALMA
imaging for eight of the twelve sources to help identify counterparts.
These eight ALMA sources in themselves form a complete sample within the
original $7'\times7'$ of the original WHDF (see Figs. 2, 3 of
\citealt{Heywood13}), although there are no ALMA data for the  nearby
spiral galaxy LAB-06. All 12 sources are contained in the extended
$16'\times16'$ WHDF area where there is a large amount of
multi-wavelength data. We shall fit optical and Near Infra-Red/Mid
Infra-Red (NIR/MIR) data to estimate redshift and dust absorption, FIR
and sub-millimetre data to estimate dust temperatures and masses.
Although we shall not model or fit the WHDF Chandra X-ray data, these
will also be vital for identifying several of our sources.

Therefore in Section \ref{sec:data} we describe the multi-wavelength
data now available in the WHDF. In Section \ref{sec:fitsed} we detail
our SED fitting procedures  before reporting the results of fitting AGN
and star-forming templates to the twelve individual SMG SED's in Section
\ref{sec:results}. In Section \ref{sec:discuss} we discuss how well the
AGN and star-forming template fits compare overall and the implications
for the nature of sub-millimetre  sources. Finally, in Section
\ref{sec:conclusions} we present our conclusions. 

We note that the analysis  of the  size, structure and SFR
surface brightness of the seven sources observed by ALMA will be
presented in Paper II (see \cite{behzad00} for preliminary results).
There, the analysis of the ALMA observations of four $z>6$ QSOs (Quasi-Stellar Objects or quasars)
\citep{Carnall15, Chehade18} will also be presented to complement the
interpretation of the WHDF SMGs. Finally, Paper III will discuss the FIR
properties from the Herschel HerMES Large Mode Survey (HeLMS)  of the
full population of sixteen  X-ray QSOs found by Chandra in the WHDF
\citep{mvm04,Bielby12} plus the four $z>6$ QSOs detected by ALMA.
 

\section{Data}
\label{sec:data}

The William Herschel Deep Field (WHDF) is a $7'\times7'$ sky area
(extended to $16'\times16'$)  at RA(J2000)=00h22m33.3s, Dec=+00d20m57s,
initially observed in the optical at the William Herschel 4.2-m
telescope \citep{NM01} and in the NIR at the Calar Alto 3.5-m telescope
\citep{NM06} to provide UBRIZHK photometry to $B\approx28$.
Subsequently, it was observed with deep Chandra X-ray data, Hubble Space Telescope (HST) i-band
imaging \citep{Bohm07,Ziegler09}, Spitzer IRAC Equatorial Survey (SPIES) 3.6 and 4.5 $\mu$m 
imaging \citep{Timlin16} and in the Herschel HeLMS survey at 250, 350
and 450 $\mu$m (e.g. \citealt{Wang15}). It has also been observed  with
deep LABOCA+ALMA exposures at sub-millimetre (sub-mm)  wavelengths
\citep{Bielby12} and at 8.4GHz(3.57cm) with the Karl G. Jansky Very
Large Array (VLA) \citep{Heywood13}. A summary of the fluxes in each
band for the SMGs can be found in Table \ref{table:LAB_flux}. Further
details including the flux errors, the coordinates of source
counterparts and their magnitudes and colours are given in  Tables 
\ref{table:LAB_error}, \ref{table:ALMA_mag}, \ref{table:LAB_mag} and
\ref{table:helms} in  Appendix \ref{app:a}.

Although WHDF was one of the earliest deep fields observed with
CCD detectors, clearly there are now other fields with wider and deeper coverage
in more high resolution bandpasses from HST, such as the GOODS fields
\citep{Dickinson03} and successors, eg CDFS \citep{Luo17}, the HST UDF
\citep{Beckwith06}, the COSMOS field \citep{Lefevre15}, CANDELS
\citep{Grogin11} etc. Nevertheless, for our current purpose, the
availability of high resolution $0.''09$, high S/N FIR data from ALMA is
ideal for checking for counterparts and looking for low surface
brightness features such as bars and spiral arms. Such data is still
only rarely available in the other deep fields. The availability of
75ksec Chandra X-ray data is also vital for diagnosing AGN. Again, other
fields such as CDFS have 7Ms exposures available and their deepest
exposure per pixel is effectively $\approx50\times$ that in WHDF.
Nevertheless, in searching for AGN X-ray signatures from typical bright
SMGs we shall see that our Chandra combination of exposure time,
resolution and low background are ideal. Fainter exposures are dominated
by X-ray galaxies below the usual $>10^{42}$ ergs s$^{-1}$ AGN limit. So
with  WHDF we can develop the analysis methods to see if a hypothesis of
significant AGN contribution to SMGs passes this first test that may
motivate similar analyses of datasets such as ALESS
\citep{Karim13,Hodge13} in ECDFS \citep{Weiss09} with  similar ALMA coverage
but deeper HST and Chandra data or AS2UDS \citep{Dud20} with many more
ALMA sources albeit observed at  lower resolution, together with deeper
ground-based K-band data.

\subsection{FIR and sub-mm data}
The original 870 $\mu$m survey in the WHDF was made in August 2008 and
May 2009 using the LABOCA instrument on the APEX telescope at the
Chajnantor site in Chile. The observations were fully described by
\cite{Bielby12} who compared eleven detected sub-mm sources in a
central $16'$ diameter area of the WHDF to the Chandra X-ray sources,
mainly identified as quasars \citep{mvm04}. The sub-mm sources were
detected as a complete sample down to 3.3mJy/beam. Only three sub-mm
sources were optically identified, LAB-05 and LAB-11 as X-ray absorbed
quasars at $z=1.32$ and $z=2.12$ and LAB-06 as a spiral galaxy at
$z=0.046$.

A complete, flux-limited (but excluding LAB-06) subset of the LABOCA
sources, LAB-01,-02,-03,-04,-05,-10, -11, in the central $7'\times7'$ of
the WHDF were then targetted  by ALMA on 11/10/2016 with the 12m Array
in a configuration which yielded 870 $\mu$m continuum images at
$0.''095$ resolution and a maximum recovered scale of $0.''926$. The
exposure times were 1572s each, long enough to detect any diffuse
emission including spiral arms and bars surrounding the sub-mm core (see
Paper II). These observations reached an 870 $\mu$m surface brightness
rms of 65 $\mu$Jy/beam over a $\approx17''$ field-of-view. All seven
targets were strongly detected with LAB-11 revealing a second sub-mm
source (named LAB-12) at $\approx5''$ from the main LAB-11 source. The
ALMA data were reduced using CASA imaging pipeline and then fitted
for ellipticity, flux and position using the IMFIT package. Full details
will be given in Paper II.

The Herschel HeLMS survey \citep{Asboth16,Oliver12}, covers the WHDF
within its full 302 deg$^2$ area. HeLMS was observed  in 2 fast-mode
scans  by Herschel SPIRE in the 250, 350 and 500 $\mu$m bands, reaching
almost the confusion limits. The resolution was $18''$, $25''$, $36''$
FWHM with $6''$, $8.''33$ and $12''$ pixels, resampled to $6''$ in the
final maps. The nominal $1\sigma$ noise (instrument+confusion) limits
are 15.61, 12.88 and  10.45 mJy in the 250, 350 and 450 $\mu$m bands.
Here we assume detection limits of $>20$mJy in each band. The 12 LABOCA
sources were cross-correlated with these Herschel catalogues and FIR
sources were taken as counterparts within radii of $20''$, $30''$,
$40''$ at 250, 350, 500$\mu$m. The resulting source fluxes are listed in
Table \ref{table:LAB_flux} and their coordinates in Table \ref{table:helms}.

\subsection{Chandra X-ray data}
The Chandra X-ray data on the WHDF turns out to be crucial in
establishing the identity of several of our sub-mm sources. Chandra
observed the WHDF with the ACIS-I detectors for 75 ksec (71 ksec
on-sky). These data and their reduction and analysis were presented by
\cite{mvm04} and an initial comparison to the WHDF LABOCA sources was
carried out by \cite{Bielby12}. The X-ray observations were made in
December 2000 with Chandra's ACIS-I detector. The observation reaches
fluxes of $4\times10^{-16}$ erg cm$^{-2}$ s$^{-1}$ in the soft 0.5-2 keV
band and $3\times10^{-15}$ erg cm$^{-2}$ s$^{-1}$ in the hard 2-8 keV
band and $1\times10^{-15}$ erg cm$^{-2}$ s$^{-1}$ in the total 0.5-8 keV
band, and resolves $>70$\% of the hard X-ray background. The ACIS-I
pointing centre was 00h 22m 33.3s, +00 20 55.0 (J2000) and the Chandra
observation id was 900079. \cite{mvm04} detected 150 sources with
$S/N>2$ in the total band. As reported by \cite{Bielby12}, only 2
sources were identified with X-rays to this limit to $S/N>3$, LAB-05,
(whdfch008) and LAB-11, (whdfch007). These X-ray sources have optical
spectrosccopy that identify them as $z=1.32$ and $z=2.12$ X-ray absorbed
QSOs.

Here, we reanalyse these X-ray data looking at the ALMA source positions
for X-ray photons to a formal $S/N>1.43$ (ie  90\% confidence) that the
source flux is non-zero in one of 3 bands, 0.5-1.2, 1.2-2.0 and 2-7 keV
using 'ciao' routine {\it srcflux}. At these limits we detected LAB-01,
LAB-03, and LAB-04 in at least one of these bands, as well as LAB-05
and LAB-11. We also detected X-rays at optical positions within the
LABOCA $\approx11''$ radius error circles for a further two sources, LAB-06 and
LAB-07. In the latter case there are two possible optical counterparts with 
X-rays, named LAB-07-1 and LAB-07-3.

We then tested for the significance of these detections by making {\it
srcflux} measurements in a randomly chosen $10\times10$ grid at
$\approx40''$ intervals in the central $7'\times7'$ ACIS-I area. No
detections were made in any band where the formal 90\% confidence
interval included zero. The faintest `sources' detected randomly had
$1.13\times10^{-5}$ c/s (ie $\approx1$ count in 71 ksec) or $8.5\times10^{-17}$ erg
cm$^{-2}$ s$^{-1}$ at 0.5-1.2 keV, $1.07\times10^{-5}$ c/s or $5.83\times10^{-17}$
erg cm$^{-2}$ s$^{-1}$ in the 1.2-2 keV band, $1.72\times10^{-5}$ c/s or
$1.07\times10^{-15}$ erg cm$^{-2}$ s$^{-1}$ in the 2-7 keV band.

We note that these latter flux limits indicate they are 99\% flux limits
when the {\it srcflux} computed flux limit was at the 90\% level. We
suggest that this is due to the background level in these early Chandra
observations being low at the level of 0.05 counts per pixel in the two
softer bands and 0.15 counts/pixel in our hard band. With ACIS-I having
$0.''492$ pixels and 50\% of light encircled within a $0.''75$ radius
within the $7'$ WHDF field,  ACIS-I  there are only $\approx7$ pixels
per resolution element giving a background count of $\approx0.35$ counts
in the softer bands and $\approx1$ count in the hard band. This means a
single count is usually enough to establish a detection in the two
softer bands and 1-2 counts in the hard band.  So LAB-01 we detect 1
count at 0.5-2keV and 2 counts at 2-7keV, in LAB-03 we detect 2 counts
at 1.2-2keV and in LAB-04 we detect 1 count at 0.5-1.2 keV and 1 count
at 1.2-2keV. Clearly, although the error on each flux will be of order
of 100\% of the signal the significance of detection in each case is
secure. This is confirmed empirically by our random flux measurements on
the real Chandra data. The X-ray fluxes are listed in Table
\ref{table:LAB_flux} and exact coordinates in Tables
\ref{table:ALMA_mag} and \ref{table:LAB_mag}.

\begin{figure*}
	\includegraphics[width=17cm]{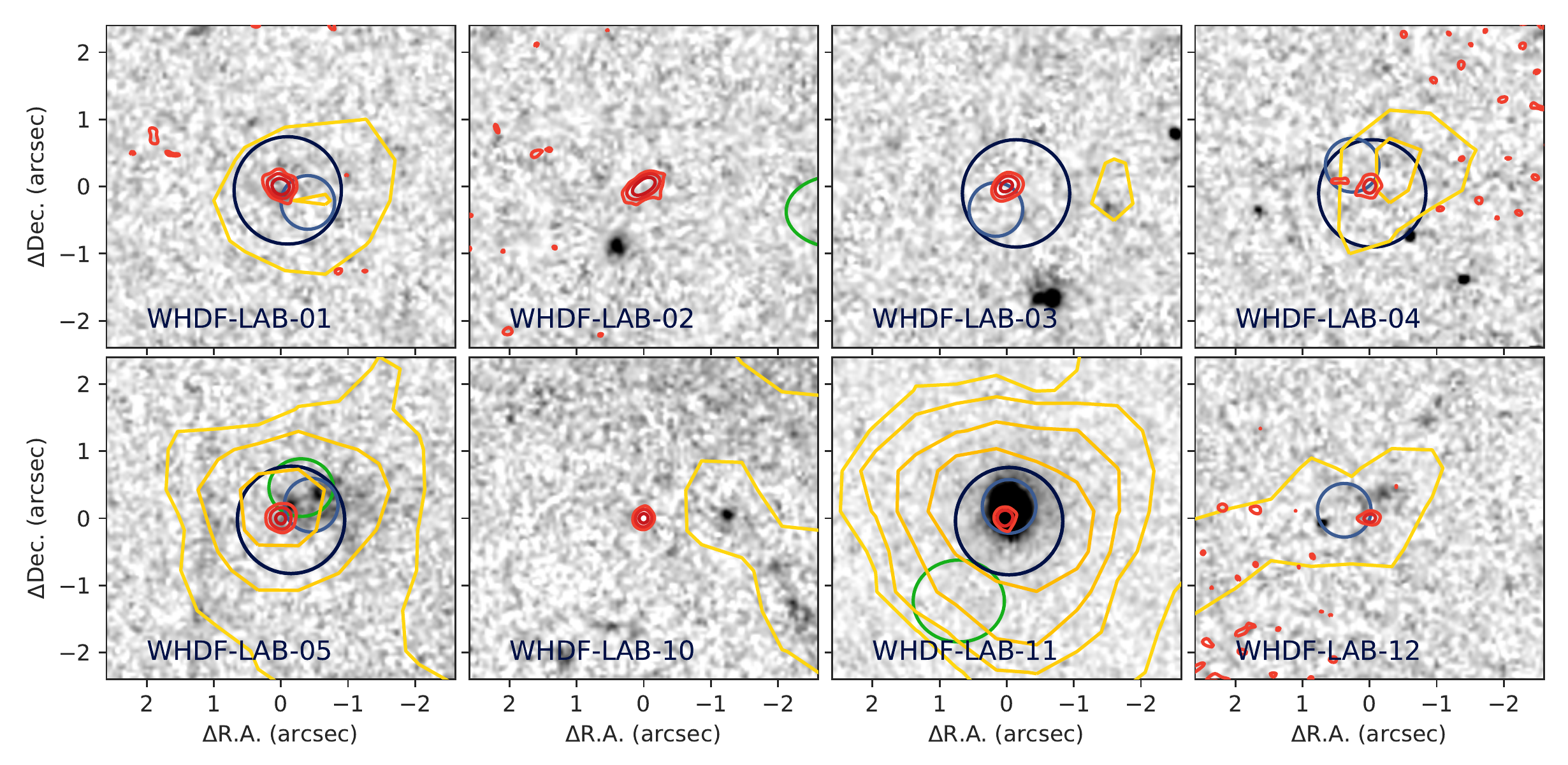}
    \caption{Finding charts (North:top, East:left) centred on 8 ALMA  sources obtained by
    combining the HST $i$ + ALMA + SPIES [3.6] $\mu$m data. The HST $i$
    image  is shown in grayscale, ALMA in red contours and SPIES in
    yellow/orange contours. Also marked are detected sources in WHDF H
    band (pale blue), Chandra X-ray (dark blue) and VLA 8.4GHz
    (green). 
    }
    \label{fig:hst_alma_spies}
\end{figure*}

\begin{figure*}
   \includegraphics[width=17cm]{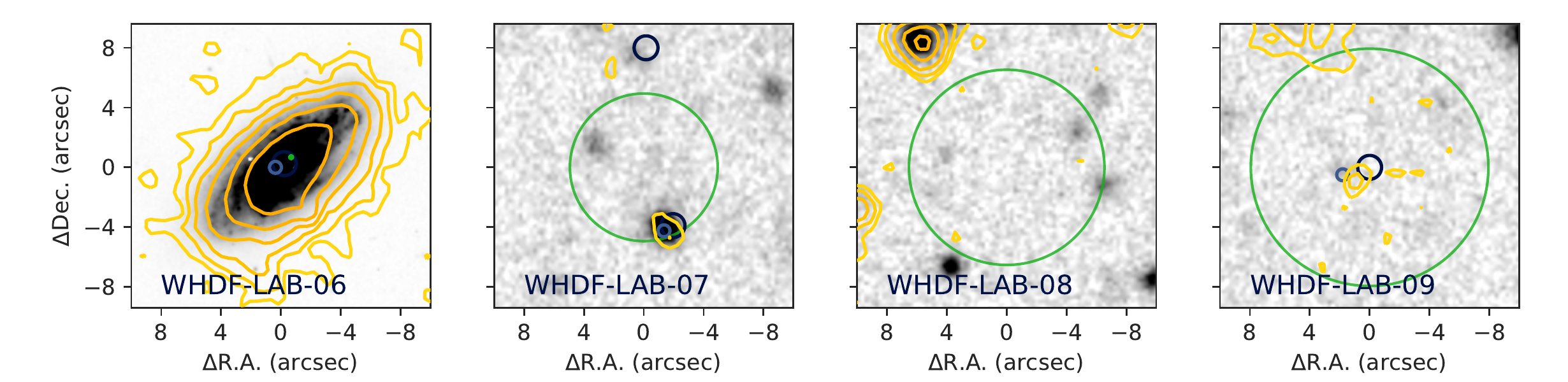}
   \caption{Finding charts (North:top, East:left) for LABOCA-only 
   sources centred on the LABOCA position in the HST $i$ band image for
   LAB-06 and the WHDF $r$ band image for the three others. The LABOCA
   error circles are shown in green. SPIES [3.6] $\mu$m data are shown
   by the yellow contours. X-ray detections are shown as black circles.
   H-band detections are shown as blue ellipses. 
   Our preferred candidate counterpart for LAB-07 (LAB-07-3) is the most Southerly
   object in the LABOCA error circle.    
   }
   \label{fig:lab6789_irband}
\end{figure*}
\begin{figure*}
   \includegraphics[width=8.5cm]{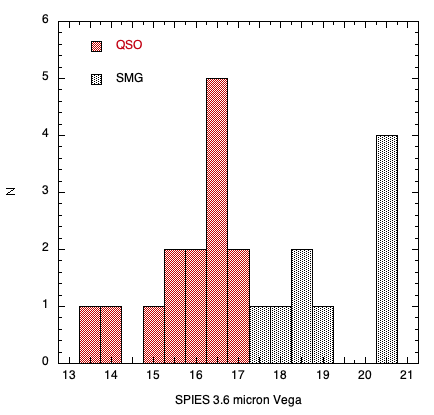}
   \includegraphics[width=8.5cm]{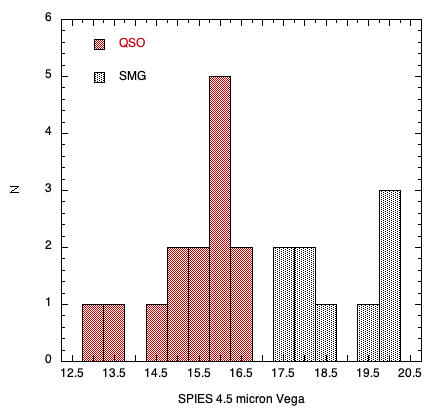}
   \caption{(a) Comparison between [3.6] $\mu$m apparent magnitudes of WHDF QSOs and SMGs, excluding LAB-06. 
   The undetected LAB-02,-03,-08,-10 SMGs are represented by upper limits in the 20.5 mag bin.
    (b) Comparison between apparent [4.5] $\mu$m apparent magnitudes of WHDF QSOs and SMGs.
    The undetected LAB-02,-03,-10 SMGs are represented by upper limits in the 19.75mag bin.
   }
   \label{fig:qso_smg_spies_mag}
\end{figure*}

\begin{figure}
    \includegraphics[width=0.5\textwidth]{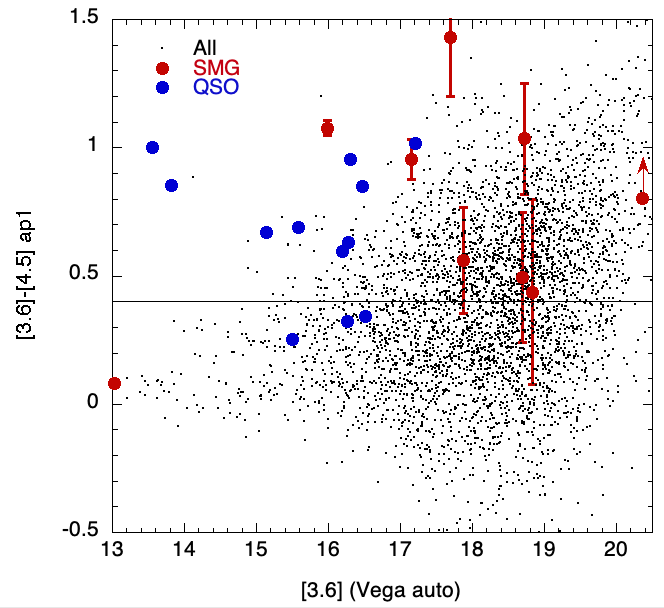}
    \caption{$[3.6]-[4.5]:[3.6]$ colour-magnitude plot for all objects
    in WHDF area from SPIES data. Red dots mark WHDF SMGs with SPIES
    detections (ie excluding LAB-2,-3,-10). LAB-08 with only a 4.5$\mu$m
    detection is shown as an upper limit in in 3.6$\mu$m and a lower
    limit in [3.6]-[4.5] $\mu$m colour. Blue dots mark WHDF QSOs with
    Chandra X-ray detections.  The horizontal line represents the
    $W1-W2>0.4$ limit recommended by \protect\cite{Stern12} for QSO selection
    and we note that all our SMGs with the exception of LAB-06 lie above this line.
    }
    \label{fig:spies}
\end{figure}

\subsection{Optical/NIR data}

To identify optical sources detected in WHDF $ubriz$ images with seeing
in the range $1.''0-1.''5$ down to $S/N=3$, we take
$\Delta\theta=2.5\times 0.6\theta (S/N)^{-1}$ \citep{Bielby12} where the
seeing $\theta=1.''5$ gives $\Delta\theta=0.''75$. Noting that all
ground-based WHDF images have $0.''4$ pixels, we have 
corrected systematics in the original WHDF astrometry by comparing WHDF
$r$ catalogue  astrometry against overlapping DECam Legacy Survey
(DECaLS) astrometry. We found that:-
RA$_{DECaLS}$=RA$_{WHDF}+0.''49\pm0.''012$,
Dec$_{DECaLS}$=Dec$_{WHDF}-0.''35\pm0.''012$. The original WHDF $H$ catalogue shows
slightly different offsets relative to DECaLS:
RA$_{DECaLS}$=RA$_H+0.''15\pm0.''0.012$,
Dec$_{DECaLS}$=Dec$_H-0.''37\pm0.''012$. The HST $i$ image offsets to
DECaLS are smaller and can be neglected. Because of the remaining
uncertainties on $r$ and $H$ WHDF coordinate offsets, we shall generally
have to be flexible in interpreting  offsets to ALMA, VLA and HST
positions. Note that in Tables \ref{table:ALMA_mag}, \ref{table:LAB_mag}
and Figs. \ref{fig:hst_alma_spies}, \ref{fig:lab6789_irband} and
throughout we shall now always use the WHDF $r$ and $H$ coordinates corrected to
the DECaLS astrometric frame.

Given these astrometric issues, here for the 8 ALMA sources we follow
\cite{Bielby12} and take $\Delta\theta<1.''3$ or $\approx3$ pixels as the
upper limit on WHDF-ALMA separations. Since the galaxy sky density at
the WHDF $3\sigma$ limit of $r=26.3$ is $\approx3\times10^5$deg$^{-2}$, this
means that in a circle of radius $1.''3$ around an ALMA source. the
expected number of optical sources is $\approx0.12$. So we only expect
contamination by $\approx1$ faint WHDF source in  our 8 ALMA source
counterparts. The WHDF coordinate corrections are less important for the
sources with only, less accurate, LABOCA positions i.e. LAB-06, -07, -08 and
-09. For these we take the error circle radii, $\Delta\theta$, listed
individually in Table 1 of \cite{Bielby12}. The resulting fluxes for the WHDF counterparts
in $ubriz$ (where available) are given in Table \ref{table:LAB_flux}.

The WHDF $H$ band data in the central $7'\times7'$ from a 14.25 hr Calar
Alto OmegaPrime exposure has $0.''9$ seeing with $0.''396$ pixels and is
particularly deep, reaching $H_{Vega}=22.9$ at $3\sigma$ \citep{NM06}.
The central WHDF $K$ band data is less deep, only reaching
$K'_{Vega}=20.7$ in similar seeing in a 0.9 hr exposure with the same
instrument \citep{NM06}. For the   HST Advanced Camera for Surveys (ACS)
2450s exposure $i$-band frame on WHDF \citep{Bohm07} the resolution is
$0.''1$ and the pixel size is $0.''05$ but we still allow a radius of
$\Delta\theta=0.''65$ to allow for astrometry systematics. Again, the
WHDF $HK$ and HST $i$ band fluxes are listed in Table
\ref{table:LAB_flux}.

\subsection{SPIES MIR data} Spitzer 3.6 and 4.5 $\mu$m data from the
Spitzer IRAC Equatorial Survey (SPIES \citealt{Timlin16}) cover
$\approx115$ deg$^2$ of SDSS Stripe 82 and therefore contains the whole
WHDF within this equatorial region. The SPIES data reach $5\sigma$
depths of $6.13\mu$Jy and $5.75\mu$Jy at 3.6 and 4.5$\mu m$,
respectively. Here, for our fits we set $\approx2\sigma$ flux limits of
$2\mu$Jy in each band. The SPIES pixel size of $0.''6$ is half of the
Spitzer IRAC pixel size due to  image dithering. The Point Spread Function (PSF) FWHM
corresponds to the `warm' IRAC  values of $1.''95$ in the 3.6 $\mu$m
detector and $2.''02$ in the 4.5 $\mu$m detector. All 3 catalogues
including the Dual detection, 3.6 $\mu$m only and 4.5 $\mu$m only were
used. 2839 matches to SPIES data were found to the 14527 $r<25.9$
objects in the $16'\times16'$ extended WHDF area. No deep, longer
wavelength data at 8, 12, 65, 100 $\mu$m are available and
this will clearly make it difficult to detect hotter temperature dust
components in our SMG SED's. The only data at longer MIR wavelengths in
the WHDF are given by the much shallower WISE data in the W3 and W4
bands at 12 and 22 $\mu$m \citep{Wright10}. Only LAB-11 is detected in
these bands. See Table \ref{table:LAB_flux} for the SMG counterpart
fluxes at 3.6 and 4.5 $\mu$m. 

\subsection{VLA radio data} VLA data for the central $7'\times7'$
at 8.4 GHz (3.57cm) were reported by \cite{Heywood13}. These radio data
were aimed at helping to identify the LABOCA source counterparts and
were made in the most compact D-configuration in March-June 2010, with a
30hr exposure time. Over most of the central $7'\times7'$ WHDF field,
the observations achieved a spatial resolution of $8''$ FWHM and an rms
background noise of $2.5\mu$Jy.  41 sources were detected at $>4\sigma$
of which 17 had primary beam corrected flux densities. LAB-02, -05, -06
and -11 were identified as radio sources with LAB-05 and LAB-11 identity
confirmed as X-ray absorbed QSOs, LAB-06 as a low redshift spiral galaxy
and LAB-02 remained optically unidentified. The VLA fluxes are again
listed as the final entries in Table \ref{table:LAB_flux}.

\subsection{Source photometry}
\label{sec:photometry}

Finding charts giving the  relative positions of these multiwavelength
data are given for the ALMA sources in Fig. \ref{fig:hst_alma_spies} and
for the LABOCA-only sources in Fig. \ref{fig:lab6789_irband}. The
resulting multi-waveband source photometry is summarised in Table \ref
{table:LAB_flux} from the X-ray to the radio. Wavelengths are given in
microns and fluxes are given in mJy. Flux upper limits are given where
there was no detection in a given band. Flux errors are given in Table
\ref{table:LAB_error}. The details of how the  counterparts were
decided upon are included in the object-by-object description of the SED
fits in Section \ref{sec:SMS_SED}.  Coordinates and other details for
source counterparts and companions are given in Tables
\ref{table:ALMA_mag}, \ref{table:LAB_mag} and \ref{table:helms}.

Again we note that the optical/NIR WHDF data on which Figs.
\ref{fig:hst_alma_spies} and \ref{fig:lab6789_irband} are based have
fewer HST bands than eg the CDFS CANDELS field where the HST depth is
also $\approx1$ mag deeper. However, we are fortunate to have the HST
F814W band available to explore the morphology of any counterpart in the
optical/UV. The CDFS field also has the advantage in having HST WFC3 H
band data at nominally $0.''15$ resolution (although with $0.''13$ pixel
size) whereas the best $H$ band resolution is $0.''9$ in the WHDF. But
again, it turns out that only 1 out of 12 submm sources  are undetected
in at least one of our bands. The WHDF ground-based depth is also
highest in the U, B bands and this makes it vital to explicitly use flux
 upper limits in these bands when sed fitting. Clearly we shall return
to exploit the advantages of ALESS and AS2UDS data to test AGN fits if
the WHDF results make this appear worthwhile.

We shall be referring to the 3.6 and 4.5 $\mu$m SPIES data throughout
the following. We therefore first show a comparison between the
distribution of 3.6 $\mu$m magnitudes (Vega-auto system)   for the sub-mm
sources and the 16 X-ray quasars listed by \cite{Bielby12} in Fig.
\ref{fig:qso_smg_spies_mag}(a) and similarly for the 4.5 $\mu$m
magnitudes in Fig. \ref{fig:qso_smg_spies_mag}(b). Note that LAB-05 and
LAB-11 that were previously identified as quasars are here counted  in
the quasars rather than SMGs. LAB-06 is also excluded from either
population on the grounds of its low redshift and hence luminosity. We
see that the broad-lined quasars are $\approx10\times$ brighter than the
SMGs and we shall discuss the implications of this result in Section
\ref{sec:discuss}. Fig. \ref{fig:spies} compares the $[3.6]-[4.5]$  (Vega
aperture 1) magnitude colours of the QSOs and sub-mm sources. Here we initially
see that all but one of our sub-mm sources show similar colours to the
quasars with $[3.6]-[4.5]>0.4$, the limit used to select AGN by \cite{Stern12}.
Again these colours will be discussed as we assess the SED
fitting results in Section \ref{sec:SMS_SED}.

\begin{figure*}
	\includegraphics[width=8.5cm]{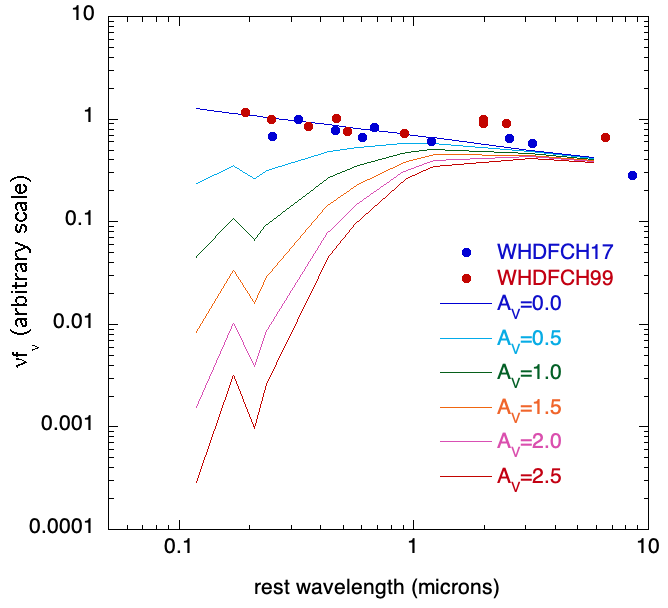}
	\includegraphics[width=8.5cm]{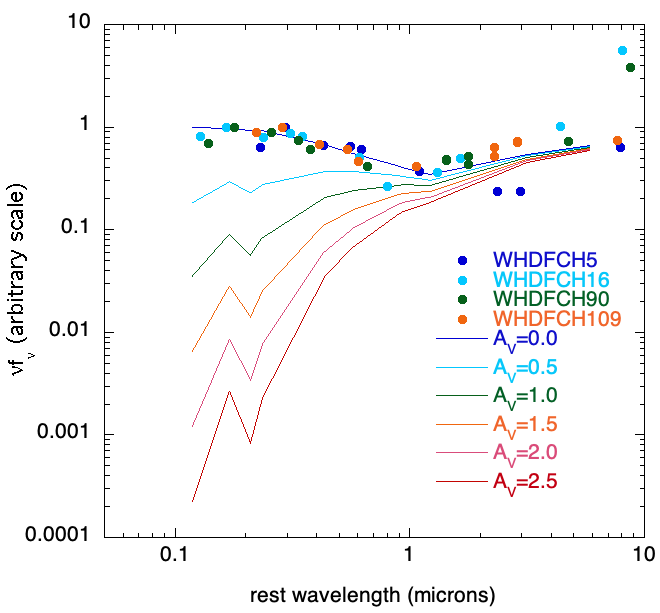}
    \caption{Single power-law and 1.4$\mu$m `Dip' fits to unobscured WHDF QSO SED's to 
     provide quasar model templates for SMG SED fitting. Also shown are the templates
     after application of our dust absorption model with $0<A_V<2.5$ mag (see Section \protect\ref{sec:dust}).
    }
    \label{fig:dip_w17_abs}
\end{figure*}

\begin{figure*}
	\includegraphics[width=8.5cm]{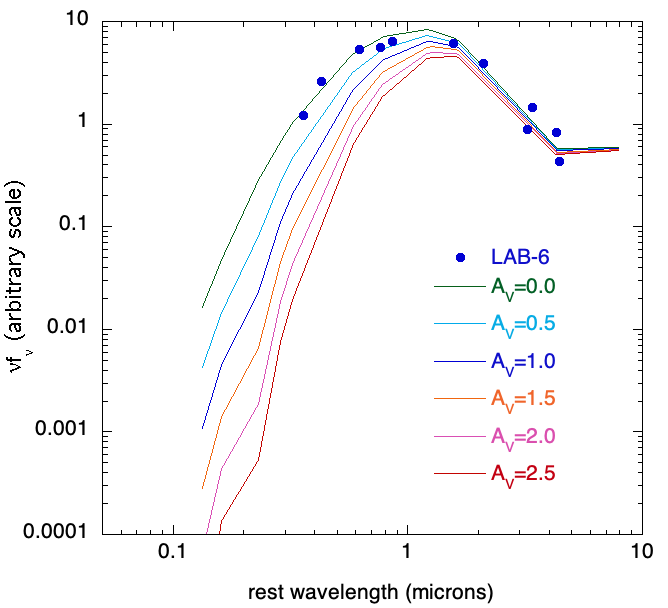}
	\includegraphics[width=8.5cm]{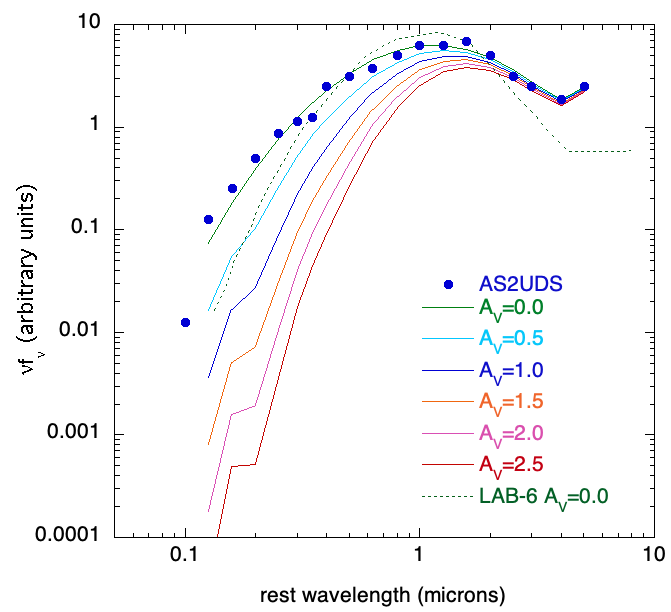}
    \caption{Fits to the LAB-06 spiral galaxy and the median AS2UDS SMG SED of
     \protect\cite{Dud20} to provide the star-forming galaxy  model
     templates for SMG SED fitting. In the right-hand panel, for
     comparison with AS2UDS, the green dotted line shows the LAB-06 template. Also
     shown are the templates after application of our dust absorption
     model with $0<A_V<2.5$ mag (see Section \protect\ref{sec:dust}).
    }
    \label{fig:dud_sf6_abs}
\end{figure*}

\begin{table*}
\begin{center}
\begin{tabular}{cccccccccccccc}
\hline
$\lambda(\mu m)$&LAB-1  &    LAB-2      &    LAB-3     &    LAB-4     &    LAB-5     &    LAB-6   &    LAB-7      &    LAB-8    & LAB-9      &    LAB-10   &    LAB-11  &    LAB-12  \\
\hline   
3.26e-4   &    6.13e-8   &  $<$1.9e-8   &    3.50e-9   &  $<$1.9e-8   &    1.73e-7   & $<$1.9e-8  &   $<$1.9e-8   &  $<$1.9e-8  &$<$1.9e-8   & $<$1.9e-8   &   3.98e-7  & $<$1.9e-8 \\
7.95e-4   &    3.02e-8   &  $<$3.6e-8   &    9.14e-8   &    2.89e-8   &    3.71e-7   & $<$3.6e-8  &   $<$3.6e-8   &  $<$3.6e-8  &$<$3.6e-8   & $<$3.6e-8   &   1.04e-7  & $<$3.6e-8 \\
1.14e-3   &  $<$3.2e-8   &  $<$3.2e-8   &  $<$3.2e-8   &    6.40e-8   &    6.51e-8   & $<$3.2e-8  &   $<$3.2e-8   &  $<$3.2e-8  &$<$3.2e-8   & $<$3.2e-8   & $<$3.2e-8  & $<$3.2e-8 \\
0.375     &  $<$3.4e-5   &  $<$3.4e-5   &    8.63e-5   &  $<$3.4e-5   &    4.13e-4   &    0.15    &    2.91e-4    &    2.00e-4  &$<$2.5e-4   & $<$3.4e-3   &   1.21e-3  & $<$3.4e-5 \\
0.45      &    5.45e-5   &  4.45e-5     &    2.83e-4   &  $<$2.7e-5   &    7.59e-4   &    0.39    &    6.43e-4    &    2.42e-4  &$<$1.7e-4   &    2.23e-4  &   1.53e-3  & $<$2.7e-5 \\
0.65      &    2.01e-4   &  1.50e-4     &    6.55e-4   &  $<$5.8e-5   &    1.13e-3   &    1.15    &    3.16e-3    &    1.66e-4  & 1.7e-4     &    7.73e-4  &   2.94e-3  & $<$9.1e-5 \\
0.80      &    0.72e-4   &  1.79e-4     &    6.61e-4   &  $<$7.6e-5   &    1.60e-3   &    1.50    &    4.70e-3    &    5.25e-4  &$<$5.25e-4  &    $-$      &   6.08e-3  &  1.77e-4  \\
0.90      &     $-$      &     $-$      &     $-$      &     $-$      &     $-$      &    1.94    &    4.06e-3    &    $-$      &   $-$      &    $-$      &   7.52e-3  &   $-$     \\
1.65      &    2.21e-3   &  $<$7.7e-4   &    1.89e-3   &    1.92e-3   &    1.06e-2   &    3.37    &    9.55e-3    &  $<$1.9e-3  & 3.84e-3    &  $<$7.7e-2  &   3.02e-2  &  3.87e-3  \\
2.20      &     $-$      &     $-$      &     $-$      &     $-$      &    1.29e-2   &    2.83    &       $-$     &    $-$      &   $-$      &    $-$      &   4.21e-2  &  7.45e-3  \\
3.37      &     $-$      &     $-$      &     $-$      &     $-$      &     $-$      &    1.00    &       $-$     &    $-$      &   $-$      &    $-$      &   8.60e-2  &   $-$     \\
3.55      &    2.00e-2   &  $<$2.0e-3   &   2.93e-3    &    9.15e-3   &    3.88e-2   &    1.72    &    9.36e-3    &  $<$2.0e-3  & 8.21e-3    &   1.14e-2   &     0.11   &  2.36e-2  \\
4.49      &    1.42e-2   &  $<$2.0e-3   &  $<$2.0e-3   &    2.04e-2   &    4.87e-2   &    1.23    &    6.56e-3    &    3.16e-3  & 9.01e-3    &   1.82e-2   &     0.16   &  2.15e-2  \\
4.62      &     $-$      &     $-$      &     $-$      &     $-$      &     $-$      &    0.67    &       $-$     &    $-$      &   $-$      &    $-$      &     0.20   &    $-$    \\
12.01     &     $-$      &     $-$      &     $-$      &     $-$      &     $-$      &    3.70    &       $-$     &    $-$      &   $-$      &    $-$      &     0.75   &    $-$    \\
22.19     &     $-$      &     $-$      &     $-$      &     $-$      &     $-$      &    4.77    &       $-$     &    $-$      &   $-$      &    $-$      &     4.04   &    $-$    \\
250       &    40.67     &    20.60     &   $<$20      &   $<$20      &    32.52     &  129.77    &    91.10      &  42.49      & $<$20      &    22.18    &    44.0    &  $<$20    \\
350       &    32.07     &    $<$20     &   $<$20      &   $<$20      &    46.14     &   39.59    &    72.04      &  54.0       & $<$20      &    27.74    &   $<$20    &  $<$20    \\
500       &    37.72     &    24.23     &   $<$20      &   $<$20      &    42.34     &   25.57    &    61.61      &  39.64      & $<$20      &   $<$20     &   $<$20    &  $<$20    \\
850       &    5.1       &    4.3       &    5.4       &    4.1       &    4.0       &    3.9     &     8.2       &   6.2       & 5.1        &    3.3      &     3.4    &  1.68     \\
35461     &     $-$      &   1.46e-2    &     $-$      &     $-$      &    3.75e-2   &    2.01e-1 &      $-$      &    $-$      &   $-$      &    $-$      &    4.6e-2  &  $-$       \\ 
\hline
\end{tabular}
\end{center}
\caption{ALMA/LABOCA counterpart source fluxes and upper limits in mJy. See Tables \ref{table:helms}, \ref{table:ALMA_mag} and 
\ref{table:LAB_mag} for coordinates of counterparts (and companions) to the sub-mm sources. Flux errors are available in the 
Supplemental Data. Rows 1-3 give the Chandra X-ray 0.5-1.2, 1.2-2.0, 2.0-7.0 keV fluxes, rows 4-9  the WHDF UBRIZHK data and rows 9,12,13,14 list 
the W1, W2, W3, W4 band fluxes from WISE. Rows 10, 11 give the 3.6, 3.5 $\mu$m fluxes from SPIES and  rows 15-16 list the HeLMS FIR fluxes. Row 17
gives the LABOCA sub-mm fluxes in mJy/beam except for LAB-12 where the ALMA flux is given. Row 17 gives the VLA 8.4GHz flux.
}
\label{table:LAB_flux}
\end{table*}

\section{SED fitting method}
\label{sec:fitsed}

The method we use for SED fitting is maximum likelihood, $L$. Here we
broadly follow the method set out by \cite{Sawicki12} which
deals with upper limits as well as detected fluxes, coded in
Fortran.  We note that we are minimising $ln(L)\approx\chi^2/2$ so
$1\sigma$ errors on derived parameters are approximately given where
$ln(L)=ln(L_{min})+0.5\times\Delta\chi^2(\nu)$ where $\Delta\chi^2=3.53, 4.72, 5.89$ 
for $\nu=3,4,5$ fitted parameters. When all fluxes are detected this reduces to
the usual $\chi^2$  fit. Note that when there are upper limits involved,
$ln(L)$  can numerically lie below zero, unlike for $\chi^2$. We fit the
SED scaling parameter, $S$, in the optical-IR regime separately from the
FIR-sub-mm range but add the likelihoods to get the best overall fit,
analogous to adding chi-squares when all observed fluxes are detected.
We do not SED-fit the X-ray data, which we regard as being beyond the
scope of this study. However, X-ray detections remain a valuable
diagnostic for distinguishing AGN and star-forming sources.

We first fit in the optical-NIR for absorption, $A_V$, and redshift, $z$
then we fit separately for a gray-body temperature plus redshift. More
details are given in Paper II but we essentially use eqn. 1 of
\cite{Dunne00} with dust mass opacity coefficient,
$\kappa=0.077\times(\nu/352.0e9)^\beta$ m$^2$ kg$^{-1}$, with emissivity
index, $\beta=1.8$, to estimate the dust mass, $M_d$, given a dust
temperature, $T$. For the $M_d$ estimates we assumed a cosmology with
$\Omega_\lambda=0.7$, $\Omega_m=0.3$ and $H_0=100$ km s$^{-1}$ Mpc$^{-1}$.
Then we combine the two fits to derive an overall best-fit redshift,
dust temperature and mass. This procedure usually leaves the best fit
absorption, $A_V$, unaffected.

We convert all data into $mJy$ and then into $\nu f_\nu$ for fitting. We
calculate $z=0$ model values with wavelength  and compare to observed
values for the source with each spectral band's rest wavelength 
calculated for the tested redshift. The redshift range tested was
$0<z<4.5$  at 0.1 redshift intervals. Other ranges were $0<A_V<2.5$ at
0.25mag intervals and $10<T<50K$ at 1K intervals.

Since we are fitting five parameters, $z$, $A_V$, $T$  and  scaling
factors $S_{Opt/IR}$  and $S_{FIR}$, we apply $\Delta\chi^2>5.89$  to
give our $1\sigma$ errors following \cite{Press92}. This corresponds to 
$\Delta -ln(L)>2.95$ which is required since we use the Maximum
Likelihood estimates to derive parameter errors as well as the
parameters themselves. For goodness of fit measures we calculate
$\chi^2$ for the maximum likelihood model parameters. Here, we replace
data upper limits by the upper limits themselves with errors also equal
to the upper limit values, thus effectively treating all upper limits as
$1\sigma$ detections.


\section{SED Fitting Results}
\label{sec:results}

In making a comparison between AGN and star-forming galaxy SMG fits, we
first acknowledge that \cite{Dud20} has shown that, with enough assumed
model components, it is certainly possible to get good $\chi^2$ fits to
their $\approx700$ SMGs with their star-forming galaxy templates.
\cite{daCunha15} list their model template parameters as including
stellar populations' age, SFR and metallicity plus stochastic SF bursts.
The dust reddening model includes lifetime of stellar birth clouds, and
the optical depths seen by stars younger and older than the clouds. They
also allow four  dust emission components - a Polycyclic Aromatic Hydrocarbon (PAH) template, an MIR hot
dust continuum plus warm and cold dust in thermal equilibrium. Since our
intent here is to test  if two, simple, reddened AGN templates can fit
our optical/MIR data, we shall also restrict ourselves to
using two basic star-forming galaxy templates with the same, simple,  dust
reddening model to act as reference points. Therefore, we next describe
our  AGN and star-forming galaxy (SFG) templates and our  dust reddening
model.

\subsection{AGN model SED's}
\label{sec:AGN_SED}

\noindent{\bf Single Power-Law AGN Template.} We first show the SED's of
the X-ray quasars WHDFCH017 at $z=0.40$  and WHDFCH099 at $z=0.82$ from
the lists of \cite{mvm04} and \cite{Bielby12}. Both  have accurate
Chandra X-ray positions which matches to a UVX quasar. These quasars
were chosen because they have a flat optical/IR spectrum in $\nu f(\nu)$
typical of about half the  bright, broad-lined  quasars in the WHDF.
WHDFCH017, -099 have no sub-mm detection. The single power-law
model fitted to their suitably normalised optical-MIR SED's  is:

\vspace{-0.3cm}
\begin{equation}
log(\nu f_\nu(\lambda))\propto -0.28log(\lambda)-0.70
\label{eq:1pl}
\end{equation}

\noindent as shown in  Fig. \ref{fig:dip_w17_abs}. Here, $\lambda$ is the rest wavelength in microns.

\smallskip

\noindent{\bf `1.4$\mu$m Dip' AGN Template.}
A further quasar template was fitted when it was realised that there
were other WHDF quasars for which a power-law with a `dip'  at
$\approx1\mu$m was a much better fit to unabsorbed X-ray QSO data. This
can be seen in Fig. \ref{fig:dip_w17_abs} where four WHDF QSOs,
WHDFCH005,-016, -090, -109 SED's are shown, after suitable re-scaling in
the vertical direction. This $\approx1\mu$m feature, here actually
fitted at 1.44$\mu$m, is interpreted as the break between the UV bump
due to the accretion disc and the longer wavelength hot
($T\approx300-1000K$) dust black-body components in QSO spectra
(e.g.\citealt{richards03,landt11}). It is not known why other QSOs such
as WHDFCH17/WHDFCH099 do not show this feature. This `1.4$\mu$m Dip' AGN
template is represented here by power-law fits in the ranges shown:

\vspace{-0.3cm}
\begin{equation}
log(\nu f_\nu(\lambda))\propto -0.11log(\lambda)-0.10, \,\,\,\, \,\,\,(\lambda<0.27 \mu m)
\label{eq:dip1}
\end{equation}
\vspace{-0.5cm}
\begin{equation}
log(\nu f_\nu(\lambda))\propto -0.65log(\lambda)-0.41, \,\,\,(0.27<\lambda<1.44 \mu m)
\label{eq:dip2}
\end{equation}
\vspace{-0.5cm}
\begin{equation}
log(\nu f_\nu(\lambda))\propto 0.72log(\lambda)-0.62, \,\,\,(1.44<\lambda<4.32 \mu m)
\label{eq:dip3}
\end{equation}
\vspace{-0.5cm}
\begin{equation}
log(\nu f_\nu(\lambda))\propto -0.11log(\lambda)-0.10, \,\,\,\,(\lambda>4.32 \mu m)
\label{eq:dip4}
\end{equation}

\noindent Both AGN models simply take a fixed 25\% of the flux above 0.1216 $\mu$m as representing 
the effect of the Lyman $\alpha$ forest on the SED below that wavelength.

\subsection{Star-forming Galaxy model SED's}
\label{sec:SFG_SED}

\noindent{\bf LAB-06 Starforming Galaxy Template.} Although this LABOCA
source has no ALMA data, it is such a bright ($r=16.05$ mag) spiral
galaxy ($z=0.046$) that it is the unambiguous counterpart of the sub-mm
source. The position of the galaxy is $5.''0$ away from the LABOCA
position, well within the the $11.''3$ tolerance listed in Table 1 of
\cite{Bielby12}. It is therefore a typical Sab spiral at the knee of
the galaxy luminosity function, $M^*$. The fitted SED is shown in Fig.
\ref{fig:dud_sf6_abs}. In Fig. \ref{fig:lab456} we then see that this
galaxy is detected in Herschel and VLA and in Chandra X-ray at the optical position. 
The SED is fitted by a quadratic plus a power law:
\vspace{-0.2cm}
\begin{equation}
log(\nu f_\nu(\lambda))\propto -3.27log(\lambda)^2+0.24log(\lambda)-0.92\,\, (\lambda<4.4\mu m)
\label{eq:quad}
\end{equation}
\vspace{-0.5cm}
\begin{equation}
log(\nu f_\nu(\lambda))\propto 0.29log(\lambda)-0.48  \,\,\,\,\, (\lambda>4.4\mu m)
\label{eq:aspl}
\end{equation}

\noindent{\bf AS2UDS Starforming Galaxy Template.} Our other
star-forming galaxy  model was made by taking the median star-forming
galaxy fit from the $\approx700$ sub-mm sources of \cite{Dud20} in the
AS2UDS survey. Here we fitted a quadratic plus linear model over the
full optical NIR range as shown in Fig. \ref{fig:dud_sf6_abs}. The
fit to the  \cite{Dud20} median AS2UDS $\lambda f_\lambda$ in
their Fig. 8 (left) is:

\vspace{-0.5cm}
\begin{equation}
log(\nu f_\nu(\lambda))\propto 0.79+0.27log(\lambda)-2.09log(\lambda)^2\,(\lambda<3.78\mu m)
\label{eq:quad}
\end{equation}
\vspace{-0.5cm}
\begin{equation}
log(\nu f_\nu(\lambda))\propto -0.50+1.28log(\lambda) \,\,\,\, (\lambda>3.78\mu m)
\label{eq:aspl}
\end{equation}

To this we applied the dust reddening formula described in Section
\ref{sec:dust}. As noted above, \cite{Dud20} found that with sufficient
parameters it was possible to get very good fits to their data. There
are also possible flaws to our approach in that we have only allowed
dust absorption to be applied to the median template and so the
unabsorbed template represents the bluest model that we can fit.
However, our only aim is to see how close the AGN models can get to the
quality of fit of the star-forming models and this justifies our
approach of minimising the number of model parameters for both AGN and
SFG templates. So in cases where the AGN models give poor fits, the SFG
templates can still give useful reference points to see how well these basic
models do in comparison.

\subsection{Dust Reddening Law}
\label{sec:dust}


We shall now proceed to apply a simplified dust absorption/reddening
law. The reddening law is a fit to the average Milky Way reddening curve
data ($R_V=3.1$) taken from Fig. 10 of \cite{Gordon03}, following
\cite{Cardelli89} in the range $0.3<x<11.0$ where $x=1/\lambda$ with
wavelength, $\lambda$, measured in microns. The fit consists of two
polynomials in the two wavelength sections given in eqns. \ref{eq:abs2}
and \ref{eq:abs3}. This allows us to include the 2200\AA~dust absorption
feature as found in the Milky Way. Other authors have found that
including this feature is helpful in fitting high-redshift galaxy and
AGN data (e.g. \citealt{daCunha15}).  We extrapolate this fit for
$x<0.3$ (i.e. $\lambda=3.33\mu$m) with eqn. \ref{eq:abs1} i.e. we assume
a simple $1/\lambda$ law for infrared wavelengths longer than
3.33$\mu$m, beyond the range of \cite{Cardelli89} but assuming  a slope
in $A_\lambda/A_V(x)$ similar to that found by these authors between the
$J$ and $K$ bands. Thus, in the fits shown below, $A_\lambda$  is the
absorption in magnitudes at wavelength $x=1/\lambda$  and $A_V$  is the
absorption in magnitudes in the $V$  band at $0.55\mu$m. The reddening
law fits are therefore:

\vspace{-0.5cm}
\begin{equation}
             A_\lambda =  A_V\times0.25x \, \, \, (x<0.3)
\label{eq:abs1}
\end{equation}
\vspace{-0.5cm}
\begin{equation}
             A_\lambda =  A_V(0.11 - 0.35x + 0.89x^2 - 0.31x^3 + 0.035x^4)\,(0.3<x<4.6)             
\label{eq:abs2}
\end{equation}
\vspace{-0.75cm}
\begin{equation}
            A_\lambda =  A_V(16.4 -5.28x + 0.62x^2 - 0.02x^3) \, \, \, (4.6<x<11.0).
\label{eq:abs3}
\end{equation}

We shall now proceed to apply this dust reddening law to the four
optical-IR templates to judge the goodness of fit of each and the amount
of dust reddening implied.  The dust absorbed templates are shown in
Figs. \ref{fig:dip_w17_abs} and Figs. \ref{fig:dud_sf6_abs}. Figs.
\ref{fig:lab123}, \ref{fig:lab456}, \ref{fig:lab789} and
\ref{fig:lab101112} then show the best AGN and star=forming galaxy fits
for  all twelve sub-mm sources. Eight of the sub-mm sources have
accurate ALMA positions LAB-01, -02,-03, -04, -05, -10, -11, -12 and
three have only less accurate LABOCA positions, LAB-07, -08, -09 plus
LAB-06.

\subsection{Sub-mm source SED fitting}
\label{sec:SMS_SED}

\begin{figure*}
   \includegraphics[width=17cm]{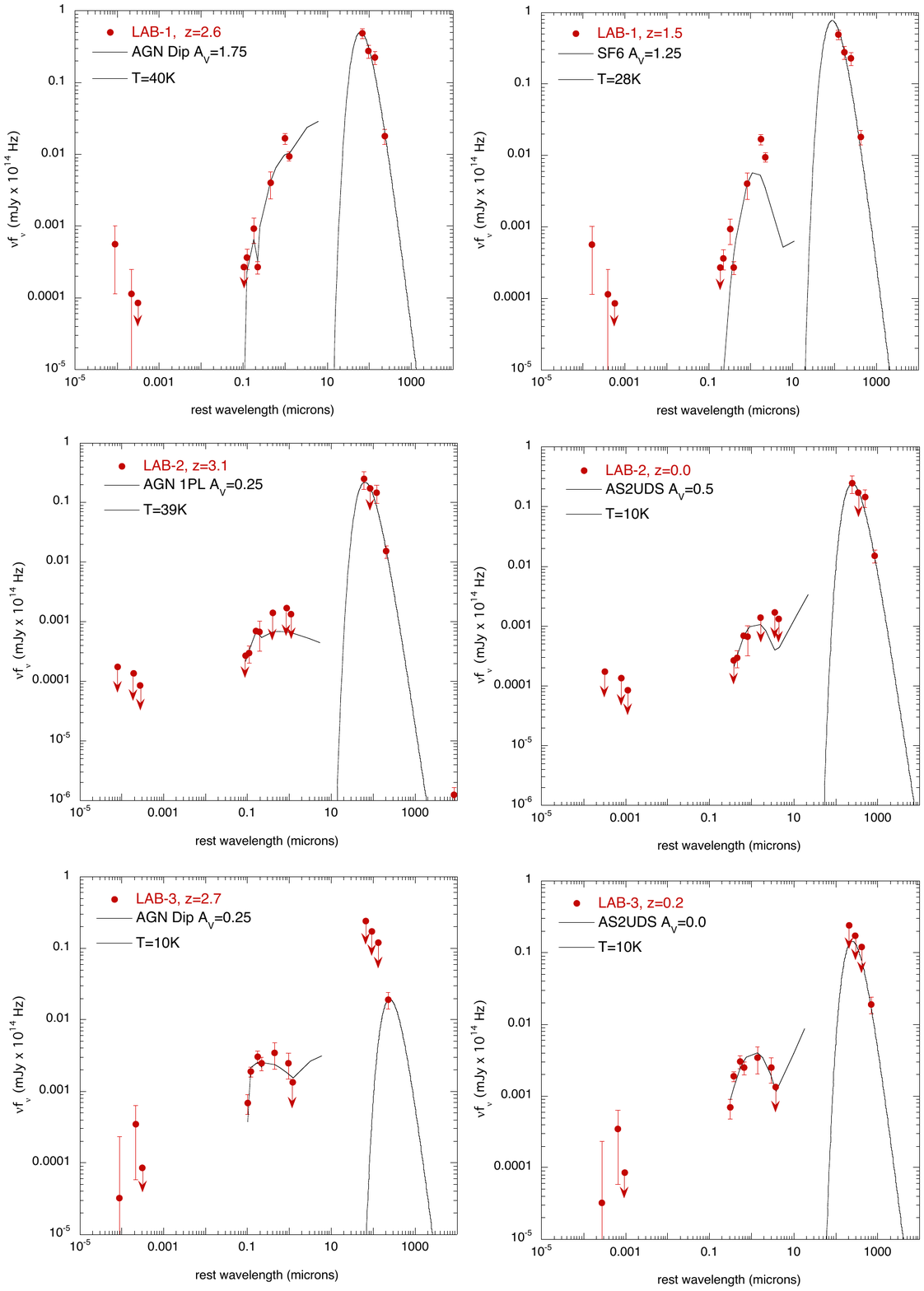}\vspace{-2cm}
   \caption{Comparison of LABOCA sub-mm sources with best fitting quasar
   and star-forming galaxy templates. Note that for LAB-03 the
   optical/MIR  detections are for a close companion since the direct
   counterpart is undetected in these bands. Also with only HeLMS FIR
   upper limits, the dust temperature (and mass) for LAB-03 are unknown and the
   best fit shown is therefore only nominal.
   }
   \label{fig:lab123}
\end{figure*}

\begin{figure*}
    \includegraphics[width=17cm]{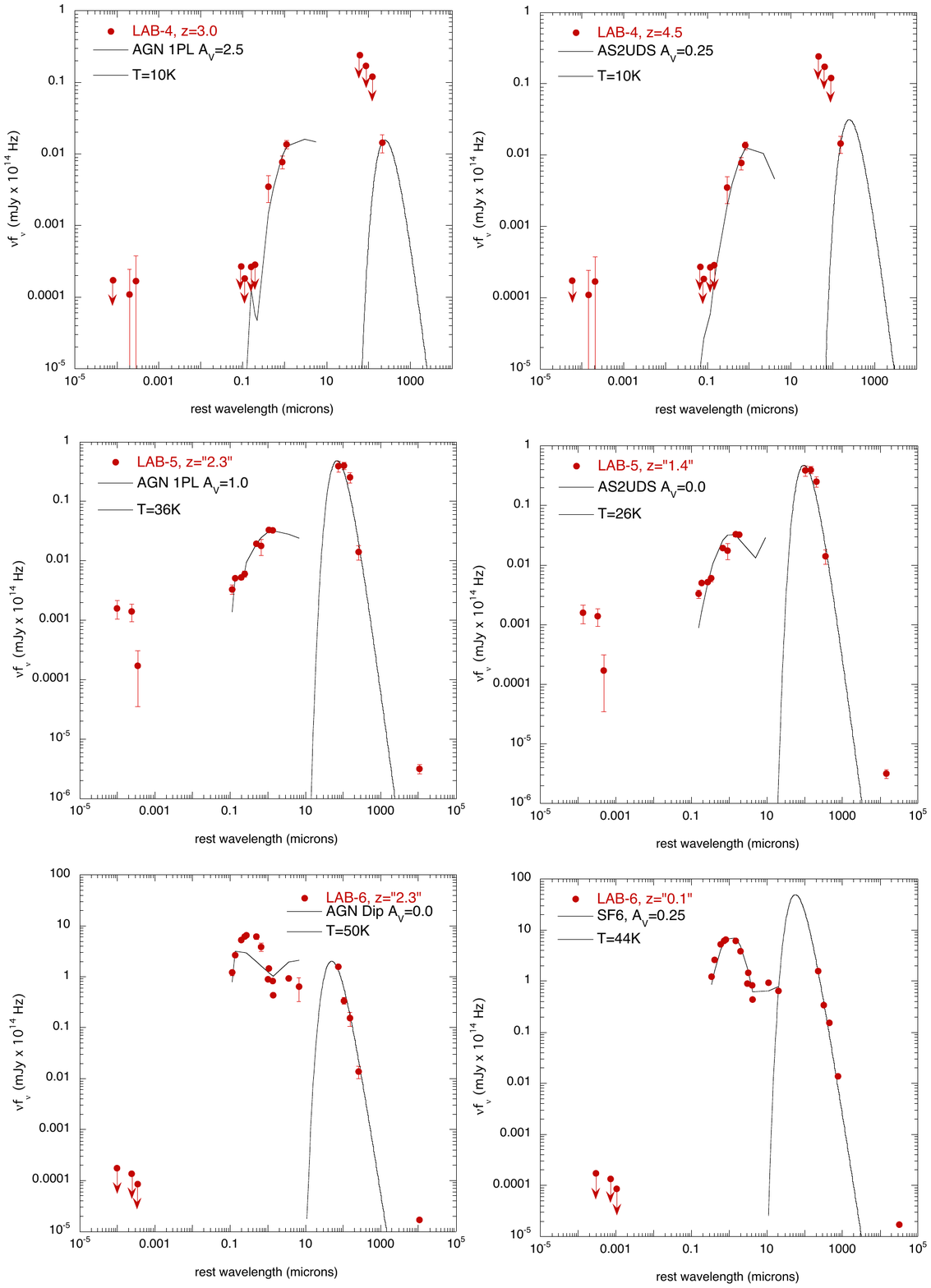}\vspace{-2cm}
    \caption{Comparison of LABOCA sub-mm sources with best fitting
    quasar and star-forming galaxy templates. Note that with only HeLMS
    FIR upper limits, the dust temperature (and mass) for LAB-04 are
    unknown and the best fit shown is therefore only nominal. Also,  for
    LAB-05 the fits are shown for the best SED fitted redshifts, whereas
    its spectroscopic redshift is $z=2.12$. The spectroscopy also
    identifies LAB-05 as a QSO \protect\citep{Bielby12}.  Similarly,
    LAB-06 has been spectroscopically identified as a $z=0.046$ spiral
    galaxy and has been used here as the basis for the star-forming
    galaxy template, SF6.
    }
    \label{fig:lab456}
\end{figure*}

\begin{figure*}
    \includegraphics[width=17cm]{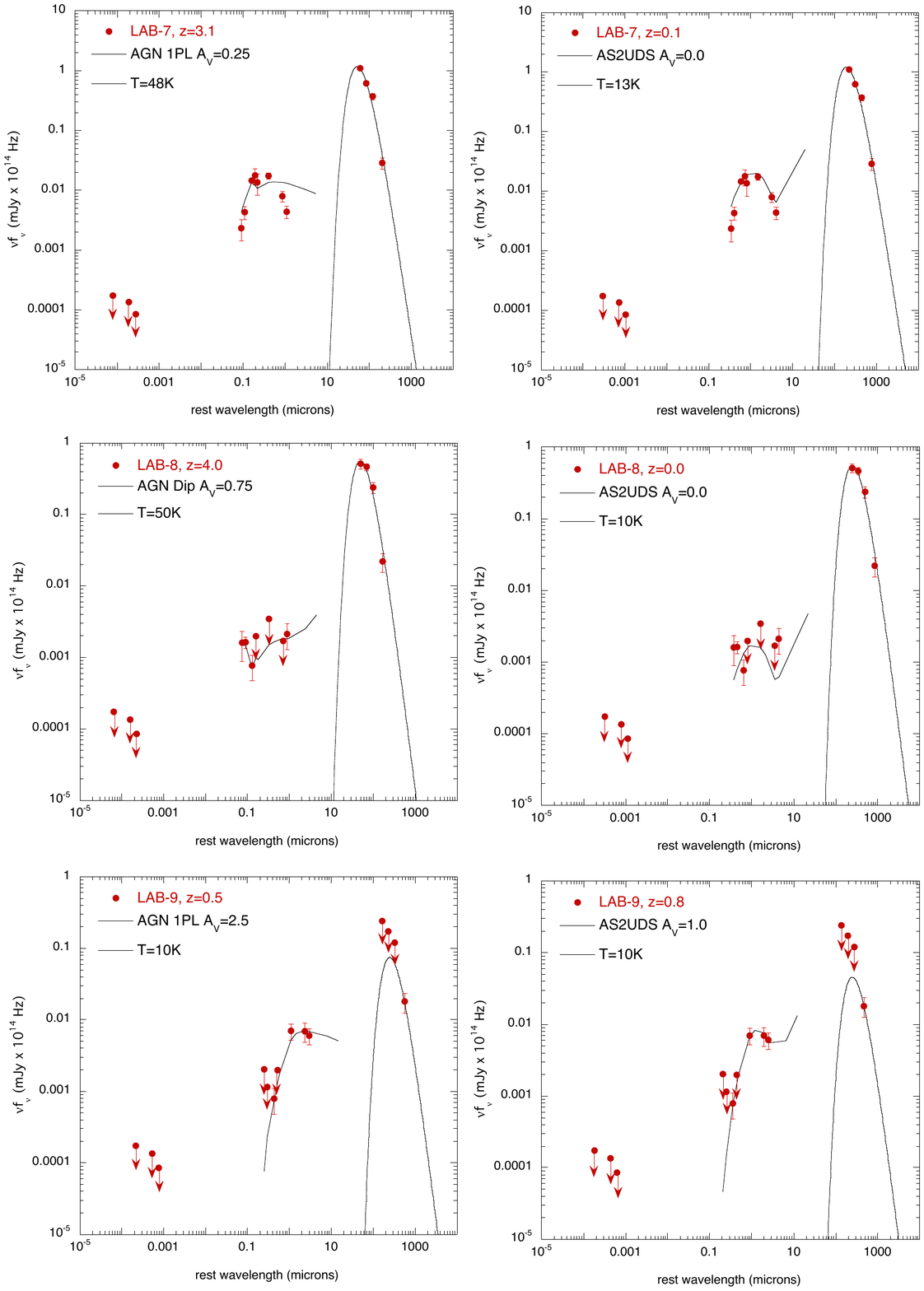}\vspace{-2cm}
    \caption{Comparison of LABOCA sub-mm sources with best fitting
    quasar and star-forming galaxy templates. Note that none of these
    three LABOCA sources have been observed by ALMA and so their sub-mm
    positions and hence  identification of  counterparts are less
    certain than for the other sources. Also, with only HeLMS FIR upper
    limits, the dust temperature (and mass) for LAB-09 are unknown and the
    best fit shown is therefore only nominal.
   }
    \label{fig:lab789}
\end{figure*}

\begin{figure*}
    \includegraphics[width=17cm]{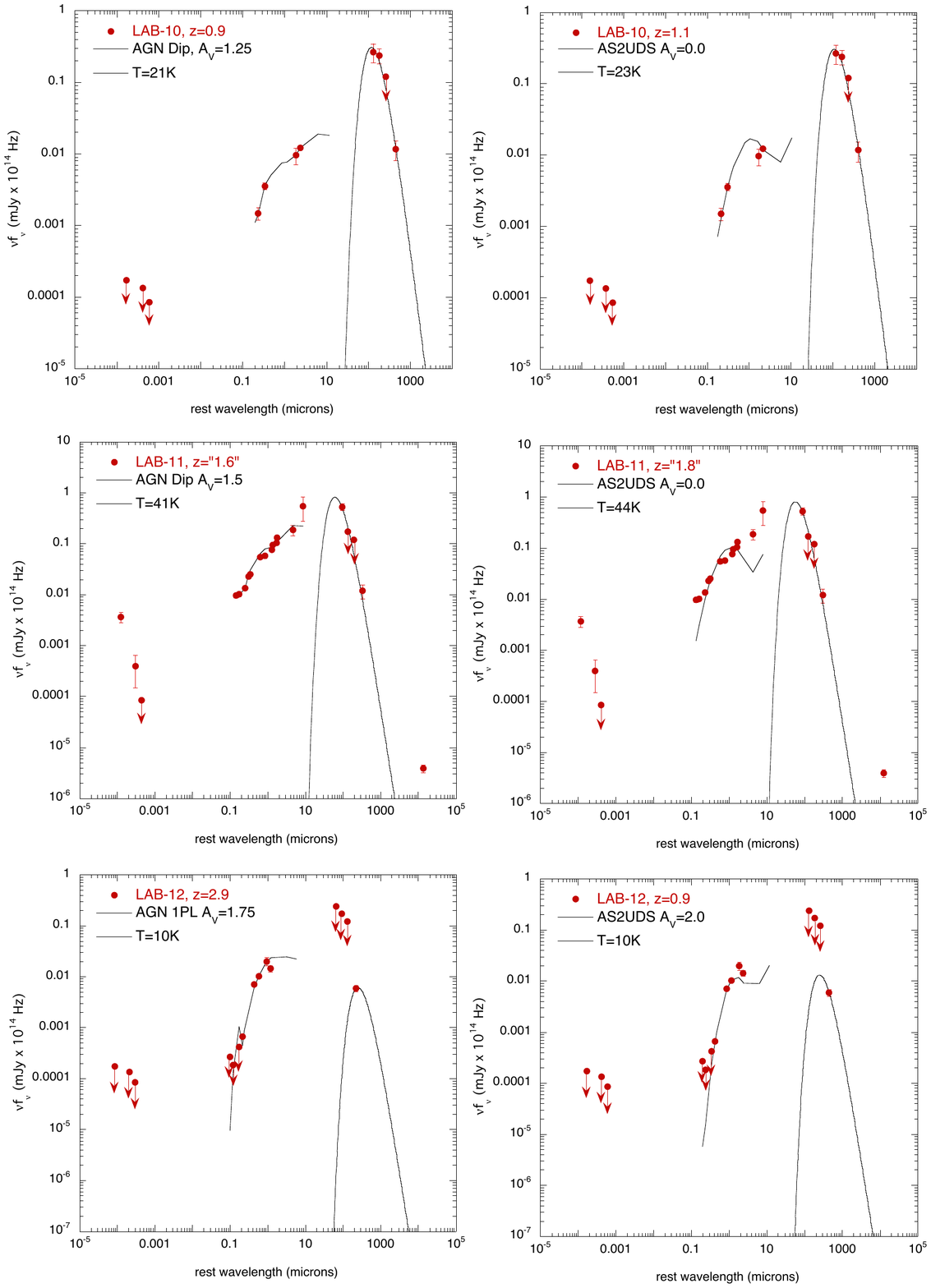}\vspace{-2cm}
    \caption{Comparison of LABOCA sub-mm sources with best fitting
    quasar and star-forming galaxy templates.  Note that  for LAB-10 the
    optical/MIR  detections are for a close companion since the direct
    counterpart is undetected in these bands. Also,  for LAB-11 the fits
    are shown for the best SED fitted redshifts, whereas its
    spectroscopic redshift is $z=1.32$. The spectroscopy also identifies
    LAB-11 as a QSO \protect\citep{Bielby12}. Finally,  with only HeLMS
    FIR upper limits, the dust temperature (and mass) for LAB-12 are
    unknown and the best fit shown is therefore only nominal.
    }
    \label{fig:lab101112}
\end{figure*}

\begin{figure}
  	\includegraphics[width=8.5cm]{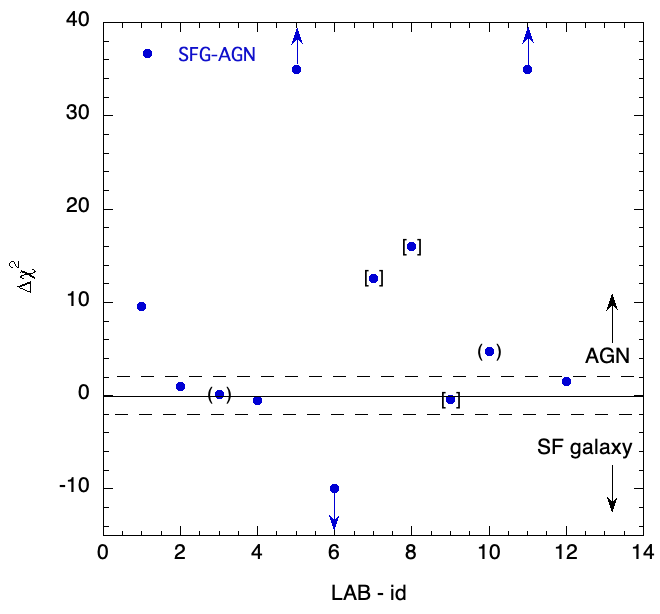}
    \caption{$\Delta\chi^2$ comparison of the best fitting star-forming
    galaxy (SFG) and  AGN models from the results in Table
    \ref{table:SED_best_fits}. LAB-03,-10 are shown in brackets since
    they are believed to be companions rather than counterparts.
    LAB-07,-08,-09 are shown in square brackets since these counterparts
    are less certain due to only having LABOCA source positions.
    }
    \label{fig:dchisq_sfg_agn}
\end{figure}

We now discuss the likely counterparts to each of our 12 sub-mm sources
and then the conclusions from the SED fits of the two AGN and two
star-forming galaxy templates. The main SED fitting results can be found
in Figs. \ref{fig:lab123}, \ref{fig:lab456}, \ref{fig:lab789},
\ref{fig:lab101112} and Table \ref{table:SED_best_fits}, with a graphical summary 
shown in Fig. \ref{fig:dchisq_sfg_agn}.
\smallskip

\noindent\underline{\bf LAB-01} In Fig. \ref{fig:hst_alma_spies} and
Table \ref{table:ALMA_mag}) we see that LAB-01 shows weak X-ray emission
at $S_X(1.2-2keV)=5.84^{+13.5}_{-4.18}\times10^{-17}$ erg cm$^{-2}$
s$^{-1}$ and $S_X(2-7keV)=7.42^{+10.1}_{-4.58}\times10^{-16}$ erg
cm$^{-2}$ s$^{-1}$ that appears to be absorbed with a column estimated
as $N_H\approx1\times10^{22}$cm$^{-2}$.  Strong FIR emission from HeLMS
is seen (see Table \ref{table:helms}) similar to LAB-05 and LAB-11
although no radio emission is detected. There is a WHDF detection at
$r=25.4$, whdf5449, only $0.''15$ offset from the ALMA position and also
detected at $b$ and $H$. There is a nearby ($0.''15$) HST i detection at
$i_{vega}=26.36\pm0.04$mag which may be slightly extended but only at
the $\approx0.''2$  level. The HST detection is difficult to see in Fig.
\ref{fig:hst_alma_spies}.  There is also a SPIES detection at $[3.6]$ +
$[4.5]$ $\mu$m at $0.''25$ from the ALMA source. It is classified as a
galaxy in both SPIES bands with $class-star-ch1=0.002$ and
$class-star-ch2=0.32$. Therefore we take the WHDF/HST/SPIES sources to
be the ALMA counterpart. 

In Fig. \ref{fig:lab123} we see that the weak X-ray emission may be
absorbed with a neutral H column estimated as
$N_H\approx1\times10^{22}$cm$^{-2}$. The comparison with the best fit
1.4 $\mu$m dip AGN model gives a dust reddening of $A_V=1.75\pm0.25$mag
and a redshift of $z=2.6\pm0.15$. The fit to the starforming galaxy is
formally worse  at least at the $\Delta\chi^2=9.54$, $4\sigma$ level for
all $A_V$ values than the AGN fits. Nevertheless, there is a suggestion
that the 1.6 $\mu$m feature in the SF6 template may fit better. In the
SPIES data, the 3.6 $\mu$m flux corresponds to $[3.6]=17.88$ (auto,Vega)
with $[3.6]-[4.5]=0.56$ for aperture 1 with a $1.''44$ radius,
consistent with an AGN and therefore suggesting that the possible 1.6
$\mu$m feature is compatible with what is seen in QSO spectra.
Conclusion: The X-ray detection suggests it is a QSO.


\noindent\underline{\bf LAB-02} In Fig. \ref{fig:hst_alma_spies} and
Table \ref{table:ALMA_mag} we see that LAB-02 shows no X-ray emission.
Strong FIR emission is seen as well as 8.4 GHz radio emission (see Table
\ref{table:helms}), although the radio source lies at $2.''9$ from the
ALMA position. The optical position (whdf9271) is also $1.''65$ away and
faint at $r=25.8$. The WHDF list classifies it as 10, signifying a
multiple source. But it is clearly detected on the HST $i$ band image as
a point source ($\approx0.''2$ FWHM) $0.''96$ offset from whdf9271. It
is also $0.''96$ from the ALMA source and we consider this WHDF/HST
source as the ALMA counterpart. A faint, $2.77\times10^{-3}$ mJy, SPIES
4.5 $\mu$m source is detected within $1.''54$ of the optical position
and $2.''18$ from the ALMA position. The source has
$class-star-ch2=0.46$, midway between point source and extended. But we
judged $2.''18$ to be too far away to be the ALMA counterpart and so
there are only SPIES $[3.6]$, $[4.5]$ $\mu$m upper limits assumed for
this ALMA source. 

In Fig. \ref{fig:lab123} the comparison with
the best fit 1 power-law  AGN model implies a dust absorption of
$A_V=0.25\pm0.38$mag, a redshift $z=3.1\pm0.25$ and $T\approx39K$. The
best fit to a starforming galaxy is AS2UDS with $A_V=0.5\pm0.75$ and
$z=0.0\pm0.15$ with a low fitted dust temperature of $T<10K$. The
star-forming fit is marginally worse than the 1-power-law AGN model at
the $\Delta\chi^2=1.0$ level but this rises to $\Delta\chi^2>9.5$ for any
$z>0.5$ whatever the $A_V$. The FIR/radio ratio also suggests a high
redshift object, with the caveat about the ALMA/radio offset. 
Conclusion: probable QSO.

\noindent\underline{\bf LAB-03} In Fig. \ref{fig:hst_alma_spies} and
Table \ref{table:ALMA_mag} we see that LAB-03 shows some evidence of
X-ray emission in the two harder bands
$S_X(1.2-2keV)=1.77^{+2.5}_{-1.1}\times10^{-16}$ erg cm$^{-2}$ s$^{-1}$
and $S_X(2-7keV)=6.29^{+4.8}_{-1.0}\times10^{-17}$ erg cm$^{-2}$
s$^{-1}$ within $0.''17$ of the ALMA position.  We are reasonably
satisfied that the WHDF source whdf9547 at $r=24.15$ is identified with
the double galaxy shown on the HST $i$ (see Fig.
\ref{fig:hst_alma_spies}) image which is offset by $1.''82$ from the
ALMA position. A SPIES 3.6 $\mu$m source close to the detection limit is
seen at $1.''57$ offset from ALMA and $1.''24$  from  whdf1234. It is
classified as a probable galaxy in the $[3.6]$ $\mu$m SPIES band with
$class-star-ch1=0.48$. This source may possibly be associated with a
fainter object detected in HST $i$, offset $1.''24$ NW of the double
galaxy. Generally, there is significant  doubt about whether any of
these optical/NIR detections are directly associated with the ALMA
sub-mm source which is significantly offset from the WHDF (+HST) galaxy
at a separation of $\approx1.''54$ and from the SPIES source at $1.''57$
(see Fig. \ref{fig:lab123}). We note that there is a faint object
detected only in the H band which may be the actual counterpart as shown
in Fig. \ref{fig:hst_alma_spies} at the `H band' coordinate listed in
Table \ref{table:ALMA_mag}. 

We have therefore fitted the SED (see Fig. \ref{fig:lab123}) assuming
the whdf1234, HST and SPIES sources are from the double galaxy but this
object is unlikely to be  the direct counterpart of the sub-mm sources
and more likely to be a companion. The SED best fit is the 1.4 $\mu$m
dip AGN model with $z=2.7\pm0.35$ and $A_V=0.25\pm0.25$, $T<10K$ but the
AS2UDS star-forming fit  with $z=0.2$, $A_V=0.0$, $T<10K$ is only
negligibly worse ($\Delta\chi^2=0.1$). In the SPIES data, $[3.6]=19.49$
(automag Vega) and it is undetected in the $[4.5]$ band. The upper limit
on the $[4.5]$ flux gives $[3.6]-[4.5]<0.31$ for aperture 1.  Here,
because of the uncertainty of the optical/NIR counterpart we weight the
detection of the weak X-ray source at the exact ALMA position more
highly and conclude that LAB-03 is an obscured quasar possibly
associated indirectly with the object at $z\approx2.7$, again bearing in
mind the caveats about the uncertain optical/NIR counterpart. 
Conclusion: QSO.

\noindent\underline{\bf LAB-04} This is the archetypal SMG in the WHDF
with no $UBRI$ detections and only detections at $H$, $[3.6]$ and
$[4.5]$. However, the $H$ and SPIES detections are relatively strong and
are both within the $0.''25-0.''5$ range. There is also a hint of
X-ray detection in the two softer bands at
$S_X(0.5-1.2keV)=1.09^{+2.5}_{-0.78}\times10^{-16}$ erg cm$^{-2}$
s$^{-1}$ and $S_X(1.2-2keV)=2.28^{+1.3}_{-0.4}\times10^{-16}$ erg
cm$^{-2}$ s$^{-1}$. although none in the harder band. The X-ray source
is within $0."095$ of the ALMA position. In the SPIES data, LAB-04 is
detected in both bands with $[3.6]=18.72$ (automag Vega) and with
$[3.6]-[4.5]=1.03$ Vega in ap1. It is also classified as a star in the
$[3.6]$ band with $class-star-ch1=0.62$ and as a galaxy in the $[4.5]$
band with $class-star-ch2=0.21$. 

From the SED's in Fig. \ref{fig:lab456} it is difficult to rule out
either an AGN or SF origin for LAB-04 although there is a marginal
$\Delta\chi^2=0.47$ preference for a $z=4.5$, $A_V=0.25$  AS2UDS
star-forming galaxy template over the $z=3.0$, $A_V=2.5$ 1-power-law
AGN. However, the $[3.6]-[4.5]=1.03$ colour satisfies the 
$[3.6]-[4.5]>0.4$ criterion for a quasar. We recall that even with
$A_V=10$ mag, $A_{3.6}=1.53$, $A_{4.5}=1.22$ the colour would only be
affected by 0.31 mag, keeping the colour well into the QSO range. On the
basis of the X-ray detection and SPIES colour: 
Conclusion: QSO.


\noindent\underline{\bf LAB-05} LAB-05 has been previously identified
showing strong X-ray emission
($S_X(0.5-1.2keV)=1.11^{+1.5}_{-0.68}\times10^{-16}$ erg cm$^{-2}$
s$^{-1}$ , $S_X(1.2-2keV)=7.18^{+3.4}_{-2.4}\times10^{-16}$ erg
cm$^{-2}$ s$^{-1}$ and $S_X(2-7keV)=2.09^{+1.1}_{-0.74}\times10^{-15}$
erg cm$^{-2}$ s$^{-1}$) that appears to be absorbed with a column
estimated as $N_H\approx1\times10^{23}$cm$^{-2}$ (\cite{Bielby12}
following \cite{mvm04}). This is confirmed with the X-ray source being
only offset from the ALMA position by $0.''38$. Strong FIR emission is
seen as well as 8.4 GHz radio emission, although the radio source EVLA-8
is offset by $1.''40$ from the ALMA position. The closest WHDF
counterpart is whdf6483, offset from ALMA by $0.''56$ (see Fig.
\ref{fig:hst_alma_spies} and Table \ref{table:ALMA_mag}) and is
classified as a galaxy in the WHDF $r$-band. This object is UVX and has
spectroscopic confirmation as a $z=2.12$ narrow-lined QSO, WHDFCH008
from \cite{Bielby12}. Now the offset between whdf6483 and the X-ray
position is $0.''65$ but we take these to be the same object and the
counterpart to the LAB-05 ALMA source. There is also a SPIES $[3.6]$,
$[4.5]$ detection at $0.''31$ offset from the ALMA source (see Fig.
\ref{fig:hst_alma_spies}). It is also classified as a galaxy in both SPIES
bands with $class-star-ch1=0.23$ and $class-star-ch2=0.07$.

The SED comparison with the best-fit single power-law AGN model  in Fig.
\ref{fig:lab456} gives a dust absorption of $A_V=1.0\pm0.25$mag. The
comparison with the best fit  1-power law AGN model gives a dust
absorption of $A_V\approx1.25\pm0.1$mag. Both give $z=2.3\pm0.15$,
consistent with the actual redshift of $z=2.12$. We note that the
reduced $\chi^2$ of the overall best fit 1.4 $\mu$m dip AGN model is
4.04 mainly due to a large residual on the $u$-band data at rest
wavelength of 0.1 $\mu$m. We decided against any increase of the
empirical error to provide a smaller $\chi^2$ so we shall bear in mind
that the AGN models are not perfect fits to this known QSO. The fit to
the starforming galaxy is very significantly worse for all $A_V$ values
(see Table \ref{table:SED_best_fits} and Fig. \ref{fig:dchisq_sfg_agn}).
We further note that the AGN absorption of  $A_V=1.25$ can be compared
to $A_V\approx 45$ mag implied by the X-ray absorption suggesting that
the X-ray absorption is located at closer line-of-sight distance to the
QSO where dust may be destroyed, with the dust that we measure via the
SED lying further away. In the SPIES data, $[3.6]=17.15$  (Vega) with
$[3.6]-[4.5]=0.96$ for aperture-1.
Conclusion: $z=2.12$  QSO.

\noindent\underline{\bf LAB-06} Although this LABOCA source has no ALMA
data, it is such a bright ($r=16.07$ mag) spiral galaxy ($z=0.046$) that
it is the unambiguous counterpart of the sub-mm sources. The position of
the galaxy (whdf3406) is $5.''0$ away from the LABOCA position, well within the the
$11.''3$ tolerance listed in Table 1 of \cite{Bielby12}. It is
therefore a typical Sab spiral at the knee of the galaxy luminosity
function, $M^*$. In Fig. \ref{fig:lab456} we then see that this
galaxy is detected in Herschel, VLA and  Chandra X-ray, the latter at a
faint flux of  $S_X(1.2-2keV)\approx6.8\times10^{-17}$ erg cm$^{-2}$
s$^{-1}$ at the optical position. We have already noted our  use of this
sub-mm galaxy as one of our two templates  for a dusty star-forming
galaxy.
Conclusion:$z=0.046$ spiral galaxy.




\noindent\underline{\bf LAB-07} 
Here we have LABOCA data but  no ALMA data. The uncertainty on the
position is $\Delta\theta=11.''6$ according to \cite{Bielby12}. There
are 3 possible optical sources associated with 3  SPIES sources at
$8.''0$ (LAB-07-1), $4.''5$ (LAB-07-2) and $3.''0$ (LAB-07-3) offsets
(see Fig. \ref{fig:lab6789_irband}). The faintest optical/SPIES source
at the largest offset is also a very faint ($\approx1\sigma$) X-ray
source in the two harder bands. But since this is below our X-ray
significance threshold and since the brightest optical/SPIES source
(LAB-07-3) is also the closest to the LABOCA position and also has a
low-significance X-ray detection, we consider this source (whdfext3652) as
the most likely counterpart. This source is classified morphologically
as a galaxy in SPIES and in WHDF. LAB-07 is also close to a strong
Herschel source at 250, 350 and 500 $\mu$m (see Table
\ref{table:helms}). 

In terms of SED fitting, both  AGN fits give $z=3.1\pm0.1$, 
$A_V=0.0\pm0.1$ and $T=48K$ and are significantly better than either of
the star-forming galaxy fits with SF6 giving the best fit with $z=0.1$,
$A_V=0.0$ and $T=13K$. Moreover, the AS2UDS SED fits both strongly reject
redshifts $z>0.5$. Also, the SPIES data give $[3.6]$=18.69 with 
$[3.6]-[4.5]=0.49$ for aperture-1. Again this seems not to  fit with a low
redshift galaxy. Although  there is uncertainty in the counterpart, the better 
AGN fit, low significance X-ray detection and red SPIES colour suggest a QSO.
Conclusion: Probable QSO

\noindent\underline{\bf LAB-08} Again we have LABOCA but no ALMA data.
The uncertainty on the LABOCA position is $\Delta\theta=12.''6$
according to \cite{Bielby12}. No X-rays nor radio emission are detected.
The optical source  whdfext3250 is faint at $r=25.6$ and lies at $4''.2$ E,
$0.''8$ N from the LABOCA coordinate (see Fig.
\ref{fig:lab6789_irband}).  Two very low S/N SPIES 3.6 $\mu$m sources
and one similarly low S/N 4.5 $\mu$m source are detected but all at
slightly larger distances ($4.''7$, $5.''8$, $7.''6$) from the LABOCA
position than whdfext3250. However, a second faint 4.5 $\mu$m  source is
detected at higher S/N closer to the whdfext3250 position  ($2.''8$ E,
$2.''07$ N). We therefore take whdfext3250 as the counterpart $4.''3$ from
the LABOCA position as this is the closest candidate, together with this
$[4.5]$ $\mu$m source, despite these larger than usual offsets between
the optical and the $[4.5]$ $\mu$m positions. Morphologically, whdfext3250
is identified as a galaxy in the $r$ band and the 4.5 $\mu$m band.
Although not detected at 3.6 $\mu$m, whdfext3250 is strongly detected in
the $u$ band and is formally UVX at $u-b=-0.67\pm0.2$. LAB-08 is also
close to a strong Herschel HeLMS source at 250, 350 and 500 $\mu$m.


The AGN SED fits are both much better than the star-forming galaxy fits
with the best 1-power law AGN model with $z=4.0\pm0.2$, $A_V=0.75\pm0.5$
nd $T>50K$ having $\Delta\chi^2=16.02$ over the best AS2UDS fit with 
$z\approx0.0$, $A_V\approx0.0$ and $T>50K$. However,  it has to be said
that both best fits are unconvincing in Fig. \ref{fig:lab789} and both
give $z=0.0$ (see Table \ref{table:SED_best_fits}). The lack of a
detection in 3.6 $\mu$m band gives a lower limit of $[3.6]-[4.5]>0.80$
which is again consistent with a QSO. This and its UVX property suggest
a QSO but there is still uncertainty over whether it is the counterpart. 
Conclusion: Probable QSO.

\noindent\underline{\bf LAB-09} Again we have LABOCA but no ALMA data.
Neither VLA radio emission nor Herschel FIR emission is detected. Only
a weak X-ray source in the softest and hardest bands may be detected at
the LABOCA position but these  are not even jointly significant at
$1\sigma$. A faint optical/IR source (whdfext5230) with $r=25.6$ and
$H=21.05$ is detected within $2''$ of the LABOCA position. whdfext5230 is
coincident with a source detected in SPIES at 3.6 and 4.5 $\mu$ms (see
Fig. \ref{fig:lab6789_irband} and Table \ref{table:LAB_mag}). The
[3.6]-[4.5]=0.44 colour is above the 0.4 mag threshold for a QSO. With
class-star-ch1=0.407 and class-star-ch2=0.117, both correspond to
galaxies and similarly for the $r$-band morphology.

The best overall SED fit is given by the AS2UDS star-forming galaxy 
model with $z=0.8\pm0.35$, $A_V=1.0\pm0.75$ and formally $T<10K$ but
with 3 upper limits from HeLMS there is essentially no constraint on the
dust temperature. However, the $\Delta\chi^2=0.43$ is only a marginal 
advantage for the AS2UDS template over the single power-law  AGN
template which also gives a low redshift ($z=0.5\pm0.35$). However, a
$z\approx0.5$ QSO is more consistent with the SPIES colour than a
$z\approx0.8$ galaxy. Although the counterpart is also uncertain, we
therefore conclude that this source is a probable QSO. 
Conclusion: Probable QSO.

\noindent\underline{\bf LAB-10} This source has ALMA data as well as
LABOCA data. We note that it lies in the outskirts of a nearby
early-type galaxy and this makes detection more difficult in the optical
{bands. Neither radio nor X-ray emission is detected. Herschel FIR
emission is detected at 250 and 350 $\mu$m. An optical source, whdf428
is detected at $r=23.97$ and $b=25.62$, offset by  $1.''68$ from the
ALMA position (see Fig. \ref{fig:hst_alma_spies} and Table
\ref{table:LAB_mag}). A point-like source is also detected on the HST
$i$ frame, $1.''2$  West of  the ALMA source at $i=25.85\pm0.05$ and
mid-way to the WHDF source with a further $1.''2$ offset and is probably
the counterpart to whdf428. A source is also detected quite strongly at
3.6 and 4.5 $\mu$m in SPIES at $1.''9$ offset from the ALMA source. This
SPIES source is probably the HST diffuse source to the SW of the HST
point source. Although the HST point source could be the counterpart of
whdf428 and the SPIES source, we regard the $1.''2$ offset of the HST
point source from the ALMA source as too far for it to be its
counterpart. Although the offsets for optical and SPIES sources are
$>1''.0$ we treat these as the detected components of a single object,
given the possible astrometric errors. We therefore treat whdf428 as a
companion to the ALMA source rather than its direct counterpart. In SED
fitting, we quickly found that the HST $i=25.85$ mag was too
faint for continuity with the brighter whdf428 and SPIES magnitudes so
this was eliminated from the analysis, implying that it may
only  be a second, closer, companion to the ALMA source.
Thus for the other whdf428+SPIES companion, we find that in both SPIES
bands, whdf428 is classed as a galaxy (class-star-1,-2=0.0) and
similarly in $r$ (SG=5). The SPIES colour is $[3.6]-[4.5]=0.92$, well
into the QSO locus.

The best SED fit for whdf428 (see Fig. \ref{fig:lab101112}) is given by
the AGN 1.4 $\mu$m `Dip' model with $z=0.9$, $A_V=1.25$ and $T=21K$ with }
$\Delta\chi^2=4.8$ compared to the best fit AS2UDS star-forming model (see
Fig. \ref{fig:dchisq_sfg_agn} and Table \ref{table:SED_best_fits}). We
conclude that although the SPIES colour suggests whdf428 is a QSO, its
$2.''3$ offset from the ALMA source makes it unlikely to be the
counterpart. The $i=25.85$ mag HST detection may still possibly be the
counterpart but that object is detected only in that band. 
Conclusion: Unknown.

\noindent\underline{\bf LAB-11} This is a known QSO, WHDFCH007, with
$z=1.32$ \citep{Bielby12}. In Fig. \ref{fig:hst_alma_spies} and Table
\ref{table:ALMA_mag}  we see that LAB-11 shows strong X-ray emission in
the two harder bands, ($S_X(1.2-2keV)=2.02^{+2.0}_{-1.1}\times10^{-16}$
erg cm$^{-2}$ s$^{-1}$ and
$S_X(2-7keV)=4.81^{+1.6}_{-1.2}\times10^{-15}$ erg cm$^{-2}$ s$^{-1}$)
that appears to be absorbed with a column estimated as
$N_H\approx5\times10^{23}$cm$^{-2}$ \citep{Bielby12}. Strong FIR emission is seen as well
as 8.4 GHz radio emission.  The closest WHDF detection, whdf6423, is
offset $0.''5$ from the ALMA source. The HST detection is offset
$0.''18$ from ALMA and $0.''60$  from whdf6423. The only discrepancy is
that the VLA detection is $1.''23$ offset from ALMA. It is classed as a
galaxy in the $[3.6]$ band with $class-star-ch1=0.21$ and a star in the
$[4.5]$ band with $class-star-ch2=0.88$.  In the SPIES data,
$[3.6]=15.99$  (automag Vega) with $[3.6]-[4.5]=1.08$ for aperture 1
with a $1.''44$ radius. 

The SED fit for the 1.4$\mu$m dip AGN model (see Fig.
\ref{fig:lab101112}) gives $A_V=1.5\pm0.1$mag and $z=1.6\pm0.1$, and
$T=32\pm5$K. The fit to the  AGN 1-power-law model is worse although
giving similar parameter values. The fits to the starforming galaxies
are very significantly worse for all $A_V$ values.
Conclusion: $z=1.32$ QSO.

\noindent\underline{\bf LAB-12} This object was picked up in the same
ALMA observation as for LAB-11. It is $\approx2\times$ fainter at
870$\mu$m than LAB-11 and lying $10.''6$ SE. It is in the direction of a
second 8.4GHz radio peak detected in EVLA-6 (see Fig. 2 of
\cite{Heywood13}), now called EVLA-6-2, but EVLA-6-2 is a further
$4.''05\pm0.''75$ E and $4.''91\pm1.''91$ S of LAB-12 (see Table
\ref{table:helms}), suggesting EVLA-6-2 is not a radio counterpart of
LAB-12. No X-rays were detected from this sub-mm source. There is also
no WHDF optical detection but a faint source is detected in the HST $i$
band at $0.''43$ offset from the ALMA source with
$i_{Vega}=25.38\pm0.04$. It is also strongly detected in the $H$, $K$
bands at $0.''32$ offset with $H=21.04\pm0.02$ and $K=19.87\pm0.02$.  In
SPIES the nearest and brightest source was first thought to be detected
only in the 4.5 $\mu$m band at $0.''24$ from the ALMA source with the
nearest 3.6 $\mu$m source $1.''68$ away. However, inspection of the 3.6 $\mu$m
image indicates that the source detected there should have the same
coordinates as the 4.5 $\mu$m sources. We then find $[3.6]=17.69$ (Vega auto)
and $[3.6]-[4.5]=1.43$ (ap1 Vega). Again this is well into the QSO
locus. 

The best SED fit is given by the AGN 1.4 $\mu$m dip model with 
$\Delta\chi^2=2.07$ over the AS2UDS star-forming template. Both models
fit reasonably well with  $\chi^2=19.27$ for the AGN model and
$\chi^2=21.34$ for the AS2UDS template with 7df.
Conclusion: Probable QSO 


\subsection{SED fitting results summary}
\label{sec:SED_results}

We have seen that the AGN models fit the SMG SED's at least as well as
the star-forming galaxy templates where the counterparts can be
identified (see Fig. \ref{fig:dchisq_sfg_agn}). The main exception is the
nearby spiral galaxy LAB-06. Of course, with a larger selection of
star-forming templates arbitrarily good fits to most SMG SED's can be
obtained as shown by \cite{Dud20}; clearly with enough model components
increasingly good fits can be expected. Here, we have therefore only
fitted two simple star-forming galaxy templates, one based on LAB-06 and
one based on the median SMG fit of \cite{Dud20}, to make a fair
comparison to the two simple AGN templates we have employed, while
keeping the number of model components low. We also caution
that the counterparts to LAB-03,-10 are unknown despite having accurate ALMA 
positions and are thus bracketted in  Fig. \ref{fig:dchisq_sfg_agn} due to 
close companions having had their SED's fitted in these two cases. Similarly, 
the counterparts to LAB-07,-08,-09 are enclosed in square brackets in Fig. 
\ref{fig:dchisq_sfg_agn} due to the uncertainty in the identity of their 
counterparts due to their only having LABOCA source positions.

One possible issue is that we have not allowed lower dust absorption than
set by the median star-forming  template so in cases where the best
star-forming fit has $A_V=0.0$, ie LAB-03, -05, -07, -08, -11 this might
bias the AGN-SF comparison. But LAB-05,-11 are previously identified
X-ray QSOs and LAB-03 also shows weak X-ray emission. So only the LABOCA
sources LAB-07,-08 are left as being possibly subject to this bias and 
none of the ALMA targets. Even LAB-07, -08 are fitted at $z\approx0$,
which is unlikely for luminous sub-mm sources. The star-forming fits
also put LAB-02 and LAB-03 at implausibly low redshifts.

A summary of the best AGN and star-forming galaxy template fits for each
sub-mm source is given in Table \ref{table:SED_best_fits}, listing  the
best-fit values for the redshift, dust temperature, $T$, dust mass,
$M_d$, and visual absorption, $A_V$. The 11 AGN redshifts (excluding
the star-forming galaxy LAB-06) range from $0<z<4.5$ with an average
$z=2.43\pm0.31$ and the absorptions range from $0<A_V<2.5$ mag with an
average of $A_V=1.25\pm0.25$ mag.  With many SFG fits being skewed to
$z=0$ we do not quote averages for these. We also do not quote average
absorption for the SFG fits because recall they do not include the extra
absorption in the original templates. For the  AGN FIR fits that have
more than one FIR data point and again excluding LAB-06 we find that the
dust mass ranges between $1.1<M_d<4.9\times10^8M_\odot$ with an average
of $M_d=1.77\pm0.46\times10^8M_\odot$. The dust temperatures for these 7
(including the LAB-08 lower limit of 50K) range from $21<T<50$K with an
average $T=39.3\pm3.6$K.  The ranges of $M_d$ for the SFG fits are much
wider and if necessary can be obtained from Table
\ref{table:SED_best_fits}. These ranges and averages for redshift,
absorption, dust temperature and dust mass are quite consistent with
those found in other sub-mm source studies where more quasar
spectroscopic redshifts are available.

\begin{table*}
\begin{center}
\begin{tabular}{cccccccccccccc}
\hline
                                                  \multicolumn{12}{c}{AGN+SFG Template Best Fits}                      \\
\hline
                           \multicolumn{7}{c}{AGN}         &           \multicolumn{5}{c}{SFG}                   \\
\hline
LAB-&$\chi^2_{AGN}$&T(K)& $A_V$(mag)   &     z      &$M_d(M_\odot)$&Templ&$\chi^2_{SFG}$&T(K) &   $A_V$(mag)  &     z       &$M_d(M_\odot)$&Templ & N \\ 
\hline
1 &    9.92&   $-$    & $1.75\pm0.25$ & $2.7\pm0.15$&   $-$    & Dip &    19.30 &   $-$   & $1.25\pm0.38$ & $1.5\pm0.15$&    $-$  & SF6 & 7  \\   
1 &   15.04& $40\pm6$ & $1.75\pm0.25$ & $2.6\pm0.15$&$2.6\times10^8$&Dip& 24.58 &$28\pm13$&$1.25\pm0.38$  & $1.5\pm0.20$&$3.0\times10^8$& SF6 & 11 \\   
2 &    1.11&   $-$    & $0.25\pm0.38$ & $3.1\pm0.25$&   $-$    & 1PL &     2.12 &    $-$  & $0.50\pm0.62$ & $0.0\pm0.15$&    $-$  & AS2 & 7  \\   
2 &    2.56& $39\pm9$ & $0.25\pm0.38$ & $3.1\pm0.25$&$2.0\times10^8$&1PL&  3.56 &   $<$10 & $0.50\pm0.75$ & $0.0\pm0.15$&$1.5\times10^4$& AS2 & 11 \\   
3 &    4.46&   $-$    & $0.25\pm0.25$ & $3.0\pm0.55$&   $-$    & Dip &     6.48 &    $-$  & $0.00\pm0.12$ & $0.1\pm0.15$&    $-$  & AS2 & 7  \\   
3 &    7.37&   n/a    & $0.25\pm0.25$ & $2.7\pm0.35$&   n/a    & Dip &     7.47 &   n/a   & $0.00\pm0.12$ & $0.2\pm0.20$&    n/a  & AS2 & 11 \\   
4 &    6.02&   $-$    & $2.50\pm0.12$ & $3.0\pm0.55$&   $-$    & 1PL &     5.52 &    $-$  & $0.25\pm1.00$ & $4.5\pm0.50$&    $-$  & AS2 & 7  \\   
4 &    8.98&   n/a    & $2.50\pm0.25$ & $3.0\pm0.60$&   n/a    &1PL  &      8.51&    n/a  & $0.25\pm1.25$ & $4.5\pm0.65$&    n/a  &AS2 & 11 \\   
5 &   15.91&   $-$    & $1.00\pm0.25$ & $2.3\pm0.10$&   $-$    & 1PL &   170.66 &    $-$  & $0.00\pm0.12$ & $1.4\pm0.15$&    $-$  & AS2 & 8  \\   
5 &   28.34& $36\pm5$ & $1.00\pm0.25$ & $2.3\pm0.15$&$3.4\times10^8$&1PL&182.80 &$26\pm4$ & $0.00\pm0.12$ & $1.4\pm0.15$&$6.5\times10^8$& AS2 & 12 \\   
6 &    9615&   $-$    & $0.00\pm0.12$ & $2.3\pm0.10$&   $-$    & Dip &     1538 &   $-$   & $0.25\pm0.25$ & $0.1\pm0.10$&    $-$  & SF6 & 13 \\   
6 &    9654& $>$50    & $0.00\pm0.12$ & $2.3\pm0.10$&$2.1\times10^8$&Dip &1543  &$44\pm5$ & $0.25\pm0.25$ & $0.1\pm0.10$&$6.7\times10^6$& SF6 & 17 \\   
7 &   18.29&   $-$    & $0.25\pm0.25$ & $3.1\pm0.10$&   $-$    & 1PL &    30.26 &    $-$  & $0.00\pm0.12$ & $0.1\pm0.10$&    $-$  & AS2 & 8  \\   
7 &   28.69& $48\pm3$ & $0.25\pm0.25$ & $3.1\pm0.10$&$3.2\times10^8 $&1PL&41.23 &$13\pm2$ & $0.00\pm0.12$ & $0.1\pm0.10$&$1.9\times10^8$& AS2 & 12 \\   
8 &    0.52&   $-$    & $0.00\pm0.25$ & $4.0\pm0.30$&   $-$    & Dip &    16.67 &    $-$  & $0.00\pm0.25$ & $0.0\pm0.10$&    $-$  & AS2 &  7 \\   
8 &    6.13&  $>$50   & $0.75\pm0.50$ & $4.0\pm0.20$&$2.2\times10^8$ &Dip&22.15 &  $<$10  & $0.00\pm0.25$ & $0.0\pm0.05$&$3.5\times10^4$& AS2 & 11 \\   
9 &    2.51&   $-$    & $2.50\pm0.25$ & $0.4\pm0.30$&   $-$    & 1PL &     1.87 &    $-$  & $1.00\pm0.62$ & $0.8\pm0.25$&    $-$  & AS2 & 7  \\   
9 &    4.17&   n/a    & $2.50\pm0.38$ & $0.5\pm0.35$&   n/a    & 1PL &     3.74 &    n/a  & $1.00\pm0.75$ & $0.8\pm0.35$&    n/a  & AS2 & 11 \\   
10&    2.03&  $-$     & $1.25\pm0.25$ & $0.9\pm0.15$&  $-$     & Dip &     6.86 &   $-$   & $0.00\pm0.12$ & $1.1\pm0.10$&    $-$  & AS2 & 7  \\   
10&    3.07& $21\pm8$ & $1.25\pm0.25$ & $0.9\pm0.20$&$4.9\times10^8$&Dip&  7.86 &$23\pm6$ & $0.00\pm0.12$ & $1.1\pm0.15$&$4.6\times10^8$&AS2& 11 \\   
11&   43.18&  $-$     & $1.50\pm0.25$ & $1.6\pm0.10$&  $-$     & Dip &    243.01&   $-$   & $0.00\pm0.12$ & $1.8\pm0.15$&    $-$  & AS2 & 13 \\   
11&   44.21& $41\pm10$& $1.50\pm0.25$ & $1.6\pm0.10$&$1.1\times10^8$&Dip &244.02& $44\pm8$& $0.00\pm0.12$ & $1.8\pm0.15$&$9.7\times10^7$& AS2 & 17 \\   
12&   16.87&  $-$     & $1.75\pm0.25$ & $2.9\pm0.10$&  $-$     & 1PL &    18.68 &   $-$   & $2.00\pm0.25$ & $0.9\pm0.10$&    $-$  & AS2 &  8 \\   
12&   19.85&  n/a     & $1.75\pm0.25$ & $2.9\pm0.10$&  n/a     & 1PL &    21.3  &   n/a   & $2.00\pm0.25$ & $0.9\pm0.10$&    n/a  & AS2 & 12 \\   
\hline
\end{tabular}
\end{center}
\caption{SED Fits with parameters and errors based on maximum likelihood
estimates. $\chi^2$ values were calculated assuming these maximum
likelihood parameters. Templ column indicates the best fitting AGN and
star-forming galaxy template. In last column, $N$ indicates the number
of data points included in fit. For a given source, first row refers to
fit in optical/NIR/MIR and second refers to the fit now also including
FIR detections. Temperatures and dust masses are marked n/a when they
would be  based only on one detected sub-mm flux with no other 
FIR fluxes except upper limits. Dust mass, $M_d$, is measured in solar
masses assuming $\Omega_m=0.3$, $\Omega_\Lambda=0.7$ and Hubble
parameter, $h=1$. X-ray and radio data are excluded from fits. 
}
\label{table:SED_best_fits}
\end{table*}

\section{Discussion}
\label{sec:discuss}

We have compared SED fits of AGN and star-forming templates to a
complete sample of twelve sub-mm sources in the WHDF over the optical-MIR and
FIR wavelengths. Chandra X-ray data was also available for several
sources but not fitted. Eight of the twelve sources originally identified by
LABOCA have high resolution, high S/N ALMA data. Of the two AGN
templates we used, we find about half are better fitted by a single
power-law model and the other half by a model with a minimum or `dip'
at 1.4 $\mu$m. This `dip' represents the split between the `blue bump' from
the accretion disc at shorter wavelengths and hot dust components at
longer wavelengths. For the star-forming templates we used an empirical
template from the $z=0.046$ spiral galaxy, LAB-06, that is one of our 12
sub-mm sources. We also used a template fitted from the median best fit
SMG SED fitted by \cite{Dud20} to $\approx700$ AS2UDS sub-mm sources.
Here we found that the AS2UDS template was the better star-forming fit to 10/12 WHDF
sub-mm sources.

In the comparison between the best AGN and star-forming template fits,
we found that the AGN template fitted better in 5 cases (LAB-01, -05,
-07, -08, -11), the star-forming template in 1 case (LAB-06), there was
no significant difference at $2\sigma$ in 4 cases (LAB-02, -04, -09,
-12) and the optical/NIR/MIR counterpart was unidentified in 2 cases
(LAB-03, -10). The main fitted parameters were the dust extinction $A_V$
in the optical-NIR, the dust temperature in the FIR and the redshift
jointly between the two. We re-iterate that there is
increased uncertainty about the counterparts of the LABOCA only sources,
LAB-07,-08,-09, given the lower precision in their sub-mm source
positions.

The X-ray data provided further important constraints. Two sub-mm
sources, LAB-5 and LAB-11, had already been identified as absorbed X-ray
quasars by \cite{Bielby12} from the $3\sigma$ X-ray catalogue of
\cite{mvm04}. With the help of the ALMA data, a re-inspection of
the X-ray data found traces of low S/N X-rays in the case of three
further sub-mm sources, LAB-01, -03 and -04. For LAB-01 and LAB-04
the fitted redshifts place their X-ray luminosity at $L_X>10^{42}$ erg
s$^{-1}$ implying that they are quasars. The detection of X-rays in the
sub-mm source LAB-03 increases the probability that it is a quasar
although we note that the doubt about its counterpart leaves its
redshift uncertain. The X-ray identifications of LAB-01, LAB-03, LAB-04
as quasars leaves seven sub-mm sources identified as quasars
(LAB-01,-03,-04,-05,-07,-08,-11), three (LAB-02,-09,-12) with no
preference between AGN and star-forming template at $2\sigma$, one low
redshift star-forming spiral (LAB-06) and two with no optical/IR/X-ray
counterpart identified (LAB-10). 

We conclude that AGN make at least as good fits to 10/12 or $83\pm26$\%
of sub-mm sources with only 1 source ($\approx8$\%) where a star-forming
galaxy was clearly the better fit, the $z=0.046$ spiral galaxy, LAB-06.
We have also noted that the majority of these 10 sources have
$W1-W2>0.4$ consistent with them being quasars. We acknowledge that
other authors have also found excellent fits to SMG SEDs with pure
star-forming templates and, given our results, we expect that models
with a mixture of AGN and star-forming templates should also provide
good fits (e.g. Wang et al. in preparation). We again emphasise that
the minimum AGN fraction of the 8
with ALMA positions comprises those with X-ray detections i.e. LAB-01,
LAB-03, LAB-04, LAB-05, LAB-11 ie 5/8 or 62.5\%. One further caveat to
be made is that we have little data at 8-100 $\mu$m between the SPIES
and Herschel/LABOCA/ALMA wavelengths. It has been previously suggested
that hot dust at observed wavelengths of  5-24 $\mu$m differentiates
quasars from other SMGs, although they are indistinguishable in the 250,
350, 500 $\mu$m band colours (e.g. Fig. 1 of \citealt{Hatz10}). We have
argued that the 3.6, 4.5, 5.5 and 8 $\mu$m colours of SMGs showing the
same colours as quasars in Fig. 2(c) of \cite{Dud20} sample of
$\approx700$ SMGs  implies that differences in the hottest dust
component may not be significant between the two populations.

\cite{Hill11a} argued that X-ray absorbed quasars may be more likely to
show FIR emission based on LAB-05, LAB-11. They built a model where more
X-ray absorbed quasars had more dust emission, that could explain
$\approx50\%$ of the sub-mm background with a further 50\% arising from faint
star-forming galaxies. However, this model is  not unified -  in the case of the
simplest models,  face-on nuclei with little HI column might be expected
to show similar amounts of FIR dust emission  as edge-on oriented quasar
discs showing  significantly larger HI column. But the Hill \& Shanks
model would imply more dust emission in the latter case. Also, such a
model may not explain the \cite{Hatz10} results unless there were
distinct AGN populations with either hot or cold dust components.  

We therefore now consider such an adjustment to the Hill \& Shanks  model
where intrinsically brighter quasars show more hot dust emission than
fainter AGN. This is supported by our finding that our X-ray quasar
sample  is,  on the average, $\approx10\times$ brighter at 3.6-4.5 $\mu$m
than our LABOCA/ALMA sub-mm sample (see Figs. \ref{fig:qso_smg_spies_mag}).
Certainly,  the small amounts of dust extinction ($A_V<2.5$mag) seen in
our SED fits are not enough to account for the sub-luminous 3.6-4.5
$\mu$m emission of SMGs. An anti-correlation between MIR and FIR
luminosities where  dust is heated and/or destroyed in the environment of
the brightest quasars could easily explain the \cite{Hatz10} result.
Such a correlation has also been seen in other X-ray selected AGN by
\cite{Ricci17}, at least in the form of X-ray quasars with high
Eddington ratios having lower X-ray absorption columns. This result can
be interpreted as, for a fixed $M_{BH}$, higher luminosity quasars have
less dusty gas due to heating by quasar outflows. 

The sub-mm counts at $>1$mJy can then be explained by intrinsically
fainter, X-ray absorbed quasars at  $L_X^*$, while the sub-mJy counts
would still be explained by modestly star-forming galaxies like the
spiral LAB-06. In the $L_X^*$ quasar population both hot
$\approx300-1000$K dust at a few parsecs radius from the accretion disc
and cooler $\approx35$K dust at $\approx1$kpc would be found. At this
latter distance, PAH features could survive the X-rays from the quasar
nucleus and so the presence of these features would no longer present a
difficulty for a model with AGN heating (e.g. \citealt{Veilleux09}).
This revised  model would still retain the main feature of the
\cite{Hill11a} model i.e. that AGN are, in the main, responsible for
both the sub-mm and X-ray backgrounds.

The fainter MIR luminosities of SMGs could still be consistent with dust
heating being more due to star-formation if their low MIR luminosity
simply indicated they were more like galaxies than AGN. However, the SMG
bolometric luminosity still reaches the quasar level when the cold dust
component is included so it seems more natural if they were  AGN
dominated.

We shall see in Paper II that the size of the ALMA cold dust sources
here are $\approx1$kpc, consistent with the prediction from the AGN
heated model. Also, at these small sizes we shall find that the SMGs are
closer to the SFR surface brightness Eddington ratio compared to the AGN
Eddington ratio.


\section{Conclusions}
\label{sec:conclusions}

We have investigated the SED's of 12 sub-mm sources in the WHDF, using
multi-wavelength data ranging from the X-ray to the radio. Our approach
has been to focus on a small but complete sample of sub-mm sources observed
with data of the highest quality. This is particularly true in the sub-mm
where our ALMA exposures on seven targets have $0.''095$ resolution combined 
with individual $\approx0.5$ hr exposures. We find that:

\noindent {\bf 1)} From an AGN versus star-forming galaxy SED fitting
comparison and/or from the presence of an X-ray detection, 7/12 sub-mm
sources are identified as quasars, 1/12 is identified as a low redshift
spiral galaxy, 3/12 show no preference between AGN and star-forming
templates, leaving 1/12 with no optical/IR/X-ray counterpart identified.
Note that 2 of the 7 quasar sub-mm sources have previous optical spectra
that identify them as quasars that also show absorbed X-ray spectra
\citep{mvm04, Bielby12} and that a further 3 of the 7 sources show X-ray
detections at fainter levels, now positively identifying 5 out of these
7 as quasars. We conclude that most  of our SMG SEDs can be as easily
fitted by an obscured AGN as an obscured star-forming galaxy with about
half now positively identified by X-rays as obscured quasars.
We re-iterate that, as shown by e.g. \cite{Dud20}, good
star-forming fits can likely be found for these SMGs. Our only
suggestion is that AGN models are frequently found to be as good fits in
simple model comparisons.

\noindent {\bf 2)}  10 out of 12 sub-mm sources show MIR colours that satisfy the $[3.6]-[4.5]>0.4$ 
quasar selection criteria of \cite{Stern12}, confirming the result of \cite{Dud20}. However, 
since $z>2$ galaxies can also occupy these MIR regions this should again only be regarded as 
a consistency check rather than a proof that they are AGN.

\noindent {\bf 3)} The SED fits for the sub-mm sources identified as AGN imply the
ranges $0.5<z<4.5$ for redshifts, $0<A_V<2.5$ mag for dust absorption,
and $10<T<50$K for the dust temperatures. We note that the dust absorption
implied for the two quasars with well-measured  X-ray absorption is an
order of magnitude higher based on their HI column,  assuming Galactic
gas-dust ratios than indicated by SED fitting. This suggests that the dust
absorbing the optical light either does not have Galactic gas-dust
ratios or is not co-located with the neutral hydrogen gas. Dust masses 
obtained from the SED fits are in the range $1-5\times10^8$M$_\odot$.

\noindent {\bf 4)} We find that the MIR brightness of sub-mm sources are typically a
factor of $\approx10$ below those of X-ray quasars although the
bolometric luminosities  of both populations are similar with the
increased sub-mm emission compensating for decreased X-ray and optical
emission. One explanation is that brighter quasars have less cold dust
and more hot dust.

In future papers we shall study the physical size of these sub-mm
sources, including  whether their sub-mm surface brightnesses are more
likely to be explained by AGN or star-forming galaxies. We shall also 
compare these results to those for luminous quasars at high redshift ($z>6$).

\section*{Acknowledgements} This paper makes use of the following ALMA
data: ADS/JAO.ALMA/2016.1.01523.S. ALMA is a partnership of ESO
(representing its member states), NSF (USA) and NINS (Japan), together
with NRC (Canada), MOST and ASIAA (Taiwan), and KASI (Republic of
Korea), in cooperation with the Republic of Chile. The Joint ALMA
Observatory is operated by ESO, AUI/NRAO and NAOJ. We thank an
anonymous  referee for their very useful comments that have improved
the quality of this paper.

\section*{Data Availability} 
The catalogue data underlying this article are available in the article and  Appendix A.
The imaging data underlying this article will be shared on reasonable request to the corresponding author.




\bibliographystyle{mnras}
\bibliography{smg1_v20} 


\appendix
\section{WHDF Flux Errors, Coordinates and Magnitudes}
\label{app:a}

Here we present further details of the WHDF data as used in the article. Table
\ref{table:LAB_error} supplies the WHDF flux errors appropriate for the
fluxes already presented in Table \ref{table:LAB_flux} in the main body
of the text. Tables \ref{table:ALMA_mag} and \ref{table:LAB_mag}
contains the RA and Declination coordinates for each ALMA counterpart or
close companion (if any). Optical data from WHDF and HST are given along with
Chandra X-ray and Spitzer 3.6, 4.5 $\mu$m fluxes. Table
\ref{table:LAB_mag} presents the equivalent data for the sources with
only LABOCA data. Finally, Table \ref{table:helms} presents HeLMS FIR
and VLA 8.4GHz radio data for all sources detected in these bands.

\begin{table*}
\begin{center}
\begin{tabular}{cccccccccccccc}
\hline
$\lambda(\mu m)$&LAB-1&    LAB-2     &    LAB-3 &    LAB-4  &    LAB-5   &   LAB-6   &    LAB-7      &    LAB-8   & LAB-9     & LAB-10  &  LAB-11 &  LAB-12  \\
\hline   
3.26e-4   &  4.89e-8  &      $-$     & 2.18e-8  &   $-$     &  6.03e-8   &   $-$     &     $-$       &    $-$     &   $-$     & $-$     & 9.37e-8 &   $-$  \\
7.95e-4   &  3.67e-8  &      $-$     & 7.60e-8  & 3.53e-08  &  1.21e-7   &   $-$     &     $-$       &    $-$     &   $-$     & $-$     & 6.47e-8 &   $-$  \\
1.14e-3   &   $-$     &      $-$     &  $-$     & 7.78e-08  &  5.20e-8   &   $-$     &     $-$       &    $-$     &   $-$     & $-$     & $-$     &   $-$  \\
0.375     &   $-$     &      $-$     & 2.69e-5  &    $-$    &  6.87e-5   &  1.77e-2  &    0.116e-3   &  8.90e-5   &   $-$     & $-$     & 1.26e-4 &   $-$  \\
0.45      &  1.71e-5  &   1.40e-5    & 4.44e-5  &    $-$    &  4.92e-5   &  4.17e-2  &    0.154e-3   &  4.55e-5   &   $-$     & 0.45e-4 & 1.60e-4 &   $-$  \\
0.65      &  7.84e-5  &   1.46e-5    & 1.29e-4  &    $-$    &  9.42e-5   &  1.10e-1  &    0.632e-4   &  6.48e-5   & 6.72e-5   & 8.34e-5 & 2.86e-4 &   $-$  \\
0.80      &  1.44e-5  &   9.35e-5    & 1.39e-4  &    $-$    &  1.90e-4   &  1.93e-2  &    0.141e-2   &    $-$     &  $-$      & $-$     & 6.38e-4 &  7.1e-6 \\
0.90      &     $-$   &      $-$     &    $-$   &    $-$    &     $-$    &  1.94e-2  &    0.160e-2   &    $-$     &  $-$      & $-$     & 7.52e-4 &   $-$   \\
1.65      &  8.96e-4  &      $-$     & 7.68e-4  &   7.8e-4  &   9.0e-4   &  0.379    &    0.124e-2   &    $-$     & 9.72e-4   & $-$     & 3.05e-3 &   $-$  \\
2.20      &    $-$    &      $-$     &     $-$  &    $-$    &   3.87e-3  &  0.509    &     $-$       &    $-$     &  $-$      & $-$     & 4.18e-3 &  1.5e-4 \\
3.37      &    $-$    &      $-$     &     $-$  &    $-$    &     $-$    &  0.052    &     $-$       &    $-$     &  $-$      & $-$     & 6.9e-3  &   $-$  \\
3.55      &  3.46e-3  &      $-$     & 1.15e-3  &   1.88e-3 &   3.39e-3  &  7.50e-3  &    1.77e-3    &    $-$     & 2.4e-3    & 2.94e-3 & 3.79e-3 &   4.2e-3 \\
4.49      &  2.15e-3  &      $-$     &    $-$   &   2.60e-3 &   2.64e-3  &  0.70e-2  &    1.55e-3    &  1.25e-3   & 2.3e-3    & 6.0e-4  & 3.39e-3 &   2.9e-3 \\
4.62      &  $-$      &      $-$     &    $-$   &   $-$     &     $-$    &  3.28e-2  &     $-$       &   $-$      & $-$       & $-$     & 0.014   &    $-$  \\
12.01     &  $-$      &      $-$     &    $-$   &   $-$     &     $-$    &   0.26    &     $-$       &   $-$      & $-$       & $-$     & 0.173   &    $-$  \\
22.19     &  $-$      &      $-$     &    $-$   &   $-$     &     $-$    &   2.38    &     $-$       &   $-$      & $-$       & $-$     & 2.00    &    $-$  \\
250       &   6.06    &      6.73    &    $-$   &   $-$     &    6.54    &   5.97    &     5.98      &    6.25    & $-$       & 6.55    & 6.74    &    $-$  \\
350       &   6.64    &      $-$     &    $-$   &   $-$     &    6.53    &   6.28    &     5.70      &    5.98    & $-$       & 6.40    & $-$     &    $-$  \\
500       &   7.92    &      7.93    &    $-$   &   $-$     &    8.54    &   7.74    &     6.84      &    7.05    & $-$       & $-$     & $-$     &    $-$  \\
850       &   1.19    &      1.02    &    1.42  &   1.11    &    1.11    &   1.08    &     1.82      &    1.82    & 1.59      & 1.03    & 1.06    &    0.24 \\
35461     &  $-$      &    4.88e-3   &   $-$    &  $-$      & 6.28e-3    &  7.55e-3  &     $-$       &    $-$     & $-$       & $-$     & 7.81e-3 &   4.6e-4 \\ 
\hline
\end{tabular}
\end{center}
\caption{ALMA/LABOCA counterpart source flux errors in mJy corresponding
to fluxes in Table \protect\ref{table:LAB_flux}. Rows 1-3 give the Chandra
X-ray 0.5-1.2, 1.2-2.0, 2.0-7.0 keV flux errors, rows 4-9  the WHDF UBRIZHK
data and rows 9,12,13,14 list the W1, W2, W3, W4 band flux errors from WISE.
Rows 10, 11 give the 3.6, 3.5 $\mu$m flux errors from SPIES and  rows 15-16
list the HeLMS FIR flux errors. Row 17 gives the LABOCA sub-mm flux errors in
mJy/beam except for LAB-12 where the ALMA flux is given. Row 17 gives
the VLA 8.4GHz flux errors.
}
\label{table:LAB_error}
\end{table*}

\begin{table*}
\begin{center}
\begin{tabular}{cllcccccccccc}
Name         &  RA(J2000)  &    Dec     & r    &Type&u-r&b-r&  r-i &  r-z  &  r-h & r-k & 3.6$\mu$m&4.5$\mu$m\\
\hline
ALMA-LAB-01(1)& 00:22:37.58& 00:19:18.32 &      &    &       &      &       &       &      &       &     &     \\
X-ray TS (2)& 00:22:37.57  & 00:19:18.3  & $-$  &5.84e-17&7.42e-16& &       &       &      &       &     &     \\
whdf5449 (3)& 00:22:37.57  & 00:19:18.35 & 25.44& 5  &   $-$ & 1.71 &  $-$  &  $-$  & 3.79 &  $-$  &     &     \\  
 H band  (4)& 00:22:37.55  & 00:19:18.12 &      &    &       &      &       &       &      &       &     &     \\ 
$i_{HST}$(5)& 00:22:37.57  & 00:19:18.19 &      &    &       &      & 26.36 &       &      &       &     &     \\
3.6, 4.5$\mu$m(6)&00:22:37.56&00:19:18.29&      &    &       &      &       &       &      &       &17.88&17.76\\
\hline
ALMA-LAB-02 & 00:22:28.44  & 00:21:47.61 &      &    &       &      &       &       &      &       &     &     \\
whdf9271    & 00:22:28.53  & 00:21:46.65 & 25.76& 10 &  $-$  & 1.61 & 0.39  &  $-$  &  $-$ &  $-$  &     &     \\ 
$i_{HST}$   & 00:22:28.47  & 00:21:46.73 &      &    &       &      & 25.49 &       &      &       &     &     \\
4.5$\mu$m   & 00:22:28.58  & 00:21:47.97 &      &    &       &      &       &       &      &       & $-$ &19.53\\ 
\hline
ALMA-LAB-03 & 00:22:45.96  & 00:18:41.17 &      &    &       &      &       &       &      &       &     &     \\
X-ray TS    & 00:22:45.95  & 00:18:41.1  &  $-$ &1.77e-16&(6.29e-17)& &     &       &      &       &     &     \\
whdf9547-Comp& 00:22:45.94 & 00:18:39.65 & 24.15& 10 & 1.62  & 1.21 & 0.20  &  $-$  &  2.33&  $-$  &     &     \\
H band      & 00:22:45.97  & 00:18:40.86 &      &    &       &      &       &       &$\approx$22.8&&     &     \\ 
$i_{HST}$-Comp& 00:22:45.91& 00:18:39.49 &      &    &       &      &double &       &      &       &     &     \\ 
3.6$\mu$m   & 00:22:45.86  & 00:18:40.95 &      &    &       &      &       &       &      &       &19.95& $-$ \\ 
\hline
ALMA LAB-04 & 00:22:29.19  & 00:20:24.79 &      &    &       &      &       &       &      &  $-$  &     &     \\
X-ray TS    & 00:22:29.19  & 00:20:24.7  &1.09e-16&5.6e-17&  & $-$  &       &       &      &       &     &     \\
H band      & 00:22:29.21  & 00:20:25.12 &      &    &       &      &       &       & 21.8 &       &     &      \\ 
3.6, 4.5$\mu$m& 00:22:29.20& 00:20:25.00 &      &    &       &      &       &       &      &       &18.71&17.36 \\
\hline
ALMA-LAB-05 & 00:22:22.87  & 00:20:13.52 &QSO&z=2.12&        &      &       &       &      &       &     &     \\
X-ray TS    & 00:22:22.86  & 00:20:13.5  &1.11-16&7.18e-16&2.09e-15&&       &       &      &       &     &     \\ 
Xray/whdfch8& 00:22:22.86  & 00:20:13.5  &      &    &       &      &       &       &      &       &     &     \\ 
whdf6483    & 00:22:22.83  & 00:20:13.05 & 23.57& 5  & -0.50 & 0.72 & 0.58  &  $-$  & 3.62 & 4.30  &     &     \\
H band      & 00:22:22.84  & 00:20:13.72 &      &    &       &      &       &       &19.95 &       &     &      \\ 
3.6$\mu$m   & 00:22:22.85  & 00:20:13.39 &      &    &       &      &       &       &      &       &17.15&16.42\\
\hline
ALMA-LAB-10 & 00:22:35.23  & 00:24:07.52 &      &    &       &      &       &       &      &       &     &     \\
HST Comp    & 00:22:35.14  & 00:24:07.59 &      &    &       &      & 25.85 &       &      &       &     &     \\
3.6, 4.5$\mu$m& 00:22:35.10& 00:24:07.24 &      &    &       &      &       &       &      &       &18.48&17.48\\
\hline
ALMA-LAB-11 & 00:22:24.84  & 00:20:11.44 &QSO   &z=1.32&     &      &       &       &      &       &     &     \\
X-ray  TS   & 00:22:24.84  & 00:20:11.4  & $-$  &2.02e-16&4.81e-15& &       &       &      &       &     &     \\ 
Xray/whdfch7& 00:22:24.84  & 00:20:11.4  &      &    &       &      &       &       &      &       &     &     \\ 
whdf6423    & 00:22:24.87  & 00:20:11.22 & 22.53&  6 &  0.37 & 1.00 &  0.99 &  1.36 & 3.72 & 4.54  &     &     \\ 
H band      & 00:22:24.84  & 00:20:11.62 &      &    &       &      &       &       &18.81 &       &     &      \\ 
$i_{HST}$   & 00:22:24.84  & 00:20:11.62 &      &    &       &      &  $''$ &       &      &       &     &     \\
3.6, 4.5$\mu$m& 00:22:24.82& 00:20:11.48 &      &    &       &      &       &       &      &       &16.00&15.14 \\   
\hline
ALMA-LAB-12:& 00:22:25.48  & 00:20:06.60 &      &    &       &      &       &       &      &       &     &     \\
H,K bands   & 00:22:25.50  & 00:20:06.72 &      &    &       &      &       &       &21.04 &19.87  &     &     \\ 
$i_{HST}$   & 00:22:25.46  & 00:20:06.95 &      &    &       &      & 25.38 &       &      &       &     &     \\
3.6$\mu$m   & 00:22:25.57  & 00:20:05.85 &      &    &       &      &       &       &      &       &17.69& $-$ \\ 
4.5$\mu$m   & 00:22:25.46  & 00:20 06.62 &      &    &       &      &       &       &      &       & $-$ &17.31 \\
\hline
\end{tabular}
\end{center}
\caption{LABOCA source coordinates and apparent
magnitudes/colours/fluxes. For each source, row (1) shows the ALMA
870$\mu$m coordinate and then any previous identfication and redshift.
Row (2) gives  coordinates of any X-ray source detected in this paper
and then the fluxes in the respective 0.5-1.2 keV, 1.2-2.0 keV and
2.0-7.0 keV bands in ergs cm$^{-2}$ s$^{-1}$. Brackets mean the X-ray
detection includes zero at 90\% confidence. Row (3) gives the WHDF $r$
band total magnitude and aperture colours where available and as
indicated at the top line of the table.  WHDF $r$ coordinates are shown
after the addition of the $+0.''49$ and $-0.''35$ offsets required to
correct them to the SDSS/ALMA/HST/VLA coordinate system. Row (4)
provides the WHDF $H$ band coordinate if object detected at $H$.
Coordinates are shown after $+0.''10$ and $-0.''38$ corrections to 
SDSS/ALMA/HST/VLA system. Row (5) indicates the detection of an HST $i$
band source, either the sub-mm counterpart or companion as indicated.
Row (6) gives the SPIES $[3.6]$ and $[4.5]$ $\mu$m coordinates and
fluxes in mJy when available. All magnitudes are in the Vega system.
}
\label{table:ALMA_mag}
\end{table*}

\begin{table*}
\begin{center}
\begin{tabular}{cllcccccccccc}
Name          & RA(J2000)     &    Dec   &Note/\,X/\,r &X/Type&X/\,u-r&b-r&  r-i&  r-z  &  r-h & r-k & 3.6$\mu$m&4.5$\mu$m\\
\hline
WHDF-LAB-06(1) & 00:22:32.09  & 00:21:24.3&z=0.046 spiral&&     &      &       &       &      &       &     &     \\
X-ray  TS (2)  & 00:22:31.85  & 00:21:27.6&(2.44e-17)&(6.82e-17)&(1.63e-16)& & &       &      &       &     &     \\
whdf3406  (3)  & 00:22:31.89  & 00:21:27.35 & 16.05&  7 &  1.61 & 1.45 &  0.59 &  0.91 & 2.36 & 2.63  &     &     \\  
H band    (4)  &\multicolumn{2}{c}{(as for whdf3406)}&&    &      &      &       &       & 13.69&       &     &     \\
3.6, 4.5$\mu$m(5)&00:22:31.84 & 00:21:27.18 &      &       &      &      &       &       &      &       &13.03&12.91\\
\hline
WHDF-LAB-07 & 00:22:48.49  & 00:16:32.8  &      &    &       &      &       &       &      &       &     &     \\ 
LAB-07-1    &              &             &      &    &       &      &       &       &      &       &     &     \\
X-ray  TS   & 00:22:48.48  & 00:16:40.8  &$-$   &(4.75e-17)&(6.57e-16)& &   &       &      &       &     &     \\
whdfext6086 & 00:22:48.52  & 00:16:40.35 & 24.68& 5  &  $-$  & 1.01 &  $-$  &  $-$  &  $-$ &  $-$  &     &     \\  
3.6, 4.5$\mu$m&00:22:48.62&00:16:40.47&      &    &       &      &       &       &      &       &18.80&18.26\\
\hline
LAB-07-2    &              &             &      &    &       &      &       &       &      &       &     &     \\
whdfext3722 & 00:22:48.71  & 00:16:34.25 & 24.38&  5 &  $-$  & 0.46 &  $-$  &  $-$  & 2.28 &  $-$  &     &     \\  
4.5$\mu$m   & 00:22:48.67  & 00:16:33.98 &      &    &       &      &       &       &      &       & $-$ &19.50\\
\hline
LAB-07-3    &              &             &      &    &       &      &       &       &      &       &     &     \\
X-ray TS    & 00:22:48.36  & 00:16 28.9  &(7.35e-17)&$-$& $-$&      &       &       &      &       &     &     \\
whdfext3652 & 00:22:48.40  & 00:16:28.55 & 22.44& 5  & 2.01  & 2.03 & 0.62  & 0.60  & 2.38 &  $-$  &     &     \\ 
H band      &\multicolumn{2}{c}{(as for whdfext3652)}& &&    &      &       &       &20.06 &       &     &     \\
3.6, 4.5$\mu$m&00:22:48.39 & 00:16:28.53 &      &    &       &      &       &       &      &       &18.69&18.59\\
\hline
WHDF-LAB-08 & 00:22:29.66  & 00:16:05.4  &      &    &       &      &       &       &      &       &     &     \\
whdfext3250 & 00:22:29.94  & 00:16:06.20 & 25.65& 5  & -0.79 & -0.12&  $-$  &  $-$  &  $-$ &  $-$  &     &     \\  
4.5$\mu$m   & 00:22:30.13  & 00:16:08.27 &      &    &       &      &       &       &      &       & $-$ &19.39 \\
\hline
WHDF-LAB-09 & 00:22:19.90  & 00:17:00.1  &      &    &       &      &       &       &      &       &     &     \\
X-ray TS    & 00:22:19.90  & 00:17:00.1  &(5.29e-17)&$-$&(8.98e-16)& &      &       &      &       &     &     \\ 
whdfext5230 & 00:22:20.02  & 00:16:59.60 & 25.61& 5  &  $-$  &  $-$ &  $-$  &  $-$  & 4.56 &  $-$  &     &     \\ 
H band      &\multicolumn{2}{c}{(as for whdfext5230)}&&&     &      &       &       &21.05 &       &     &     \\
3.6, 4.5$\mu$m&00:22:20.00 & 00:16:59.57 &      &    &       &      &       &       &      &       &18.84&18.25\\        
\hline
\end{tabular}
\end{center}
\caption{LABOCA coordinates and apparent magnitudes/colours/fluxes for
sources with no ALMA observation. For each source, row (1) shows the
original LABOCA 870$\mu$m coordinate and then any previous identfication
and redshift. Row (2) gives  coordinates of any X-ray source detected in
this paper and then the fluxes in the respective 0.5-1.2 keV, 1.2-2.0
keV and 2.0-7.0 keV bands in ergs cm$^{-2}$ s$^{-1}$. Brackets mean the
X-ray detection includes zero at 90\% confidence. Row (3) gives the WHDF
$R$ band magnitude and aperture colours where available and as indicated
at the top line of the table. WHDF $r$ coordinates that apply to both
sources and images are shown after the addition of the $+0.''49$ and
$-0.''35$ offsets required to correct them to the SDSS/ALMA/HST/VLA
coordinate system.  Row (4) provides the WHDF $H$ band coordinate
corrected by $+0.''15$ in RA and $-0.''37$ in Dec to the
SDSS/ALMA/HST/VLA system, if object detected at $H$.  Row (5) gives the
SPIES $[3.6]$ and $[4.5]$ $\mu$m coordinates and fluxes in mJy when
available. All magnitudes are in the Vega system. For LAB-07 three
possible counterparts are given with our preferred candidate being
LAB-07-3. }
\label{table:LAB_mag}
\end{table*}

\begin{table}
\begin{center}
\begin{tabular}{cllcccccccccccl}
Name+Band   & RA(J2000)    &    Dec      &flux (mJy)& \\
\hline
ALMA-LAB-01 & 00:22:37.58  & 00:19:18.32 &       & \\
250$\mu$m   & 00:22:37.61  & 00:19:11.19 & 40.67 & \\
350$\mu$m   & 00:22:37.61  & 00:19:11.65 & 32.07 & \\
500$\mu$m   & 00:22:37.24  & 00:19:24.71 & 37.72 & \\
\hline
ALMA-LAB-02 & 00:22:28.44  & 00:21:47.61 &       &\\
250$\mu$m   & 00:22:27.84  & 00:21:30.43 & 20.60 &\\
500$\mu$m   & 00:22:27.74  & 00:21:26.64 & 24.23 &\\
EVLA-32     & 00:22:28.25  & 00:21:47.24 &1.46e-2&\\
\hline
ALMA-LAB-05 & 00:22:22.87  & 00:20:13.52 &       &\\
250$\mu$m   & 00:22:22.90  & 00:20:15.61 & 32.52 &\\
350$\mu$m   & 00:22:23.46  & 00:20:12.70 & 46.14 &\\
500$\mu$m   & 00:22:23.49  & 00:20:22.87 & 42.34 &\\
EVLA-8      & 00:22:22.81  & 00:20:14.6  &3.75e-2&\\
\hline
WHDF-LAB-06 & 00:22:32.09  & 00:21:24.3  &       &\\
250$\mu$m   & 00:22:31.98  & 00:21:26.58 & 129.8 &\\
350$\mu$m   & 00:22:31.86  & 00:21:26.50 & 39.59 &\\
500$\mu$m   & 00:22:32.94  & 00:21:25.35 & 25.57 &\\
EVLA:01     & 00:22:31.86  & 00:21:27.5  &2.01e-1&\\
\hline
WHDF-LAB-07 & 00:22:48.49  & 00:16:32.8  &       &\\ 
250$\mu$m   & 00:22:47.84  & 00:16:31.83 & 91.10 &\\
350$\mu$m   & 00:22:47.94  & 00:16:29.67 & 72.04 &\\
500$\mu$m   & 00:22:47.72  & 00:16:34.98 & 61.61 &\\
\hline
WHDF-LAB-08 & 00:22:29.66  & 00:16:05.4  &       &\\
250$\mu$m   & 00:22:30.07  & 00:16:10.89 & 42.49 &\\
350$\mu$m   & 00:22:30.07  & 00:16:08.98 & 54.0  &\\
500$\mu$m   & 00:22:30.23  & 00:16:09.57 & 39.64 &\\
\hline
ALMA-LAB-10 & 00:22:35.23  & 00:24:07.52 &       &\\
250$\mu$m   & 00:22:35.59  & 00:24:09.60 & 22.18 &\\
350$\mu$m   & 00:22:35.36  & 00:24:11.00 & 27.74 &\\
\hline
ALMA-LAB-11 & 00:22:24.84  & 00:20:11.44 &       &\\
250$\mu$m   & 00:22:25.29  & 00:20:08.92 & 44.00 &\\
EVLA-6      & 00:22:24.89  & 00:20:10.21 &4.6e-2 &\\
\hline
ALMA-LAB-12:& 00:22:25.48  & 00:20:06.60 &       &\\
EVLA-6-2    & 00:22:25.75  & 00:20:01.69 &7.6e-3 &\\    
\hline
\end{tabular}
\end{center}
\caption{ALMA/LABOCA source coordinates and HeLMS+VLA  positions and fluxes. 
 Note that EVLA-6-2 is regarded as a companion to LAB-12
 rather than a counterpart.
}
\label{table:helms}
\end{table}

\bsp	
\label{lastpage}
\end{document}